\documentclass{emulateapj}

\usepackage{color}

\newcommand{\ud}{{\rm d}}
\newcommand{\gsim}{\;\rlap{\lower 2.5pt
 \hbox{$\sim$}}\raise 1.5pt\hbox{$>$}\;}
\newcommand{\lsim}{\;\rlap{\lower 2.5pt
   \hbox{$\sim$}}\raise 1.5pt\hbox{$<$}\;}
\newcommand{\be}{\begin{equation}}
\newcommand{\beq}{\begin{equation}}
\newcommand{\ba}{\begin{eqnarray}}
\newcommand{\ee}{\end{equation}}
\newcommand{\eeq}{\end{equation}}
\newcommand{\ea}{\end{eqnarray}}

\newcommand{\bea}{\begin{eqnarray}}
\newcommand{\eea}{\end{eqnarray}}
\newcommand{\bean}{\begin{eqnarray*}}
\newcommand{\eean}{\end{eqnarray*}}

\newcommand{\bx}{{\bf x}}

\newcommand{\bn}{{\bf \hat{n}}}

\newcommand{\bV}{{\bf V}}
\newcommand{\tcmb}{T_\gamma}
\newcommand{\Lyman}{${\rm Ly\alpha}$ }

\begin{document}

\title{Cosmic Reionization and the 21-cm signal: Comparison between an analytical model and a simulation} 
\author{M\'ario G. Santos$^1$, Alexandre
  Amblard$^2$, Jonathan Pritchard$^{3,4}$, Hy Trac$^{4,5}$, Renyue Cen$^5$,
  Asantha Cooray$^2$} \affil{
  $^1$CENTRA, Instituto Superior T\'ecnico, Lisboa 1049-001, Portugal\\
  $^2$Center for Cosmology, Department of Physics and Astronomy, 4186
  Frederick Reines Hall,
  University of California, Irvine, CA 92697\\
  $^3$California Institute of Technology, Mail Code 130-33, Pasadena, CA 91125\\
  $^4$Harvard-Smithsonian Center for Astrophysics, MS-51, 60 Garden St, Cambridge, MA 02138, USA\\
  $^5$Department of Astrophysical Sciences, Princeton University,
  Princeton, NJ 08544}

\begin{abstract}
We measure several properties of the reionization process and the corresponding low-frequency 21-cm signal associated with the neutral hydrogen distribution, using a large volume, high resolution simulation of cosmic reionization. The brightness temperature of the 21-cm signal is derived by post-processing this numerical simulation with a semi-analytical prescription. Our study extends to high redshifts ($z\sim 25$) where, in addition to collisional coupling, our post-processed simulations take into account the inhomogeneities in the heating of the neutral gas by X-rays and the effect of an inhomogeneous \Lyman radiation field. Unlike the well-studied case where spin temperature is assumed to be significantly greater than the temperature of the cosmic microwave background due to uniform heating of the gas by X-rays, spatial fluctuations in both the \Lyman radiation field and X-ray intensity impact predictions related to the brightness temperature at $z > 10$, during the early stages of reionization and gas heating. The statistics of the 21-cm signal from our simulation are then compared to existing analytical models in the literature and we find that these analytical models provide a reasonably accurate description of the 21-cm power spectrum at $z < 10$.  Such an agreement is useful since analytical models are better suited to quickly explore the full astrophysical and cosmological parameter space relevant for future 21-cm surveys.  We find, nevertheless, non-negligible differences that can be attributed to differences in the inhomogeneous X-ray heating and \Lyman coupling at $z > 10$ and, with upcoming interferometric data, these differences in return can provide a way to better understand the astrophysical processes during reionization.
\end{abstract}

\keywords{cosmology: theory --- large scale structure of
Universe --- diffuse radiation}

\submitted{Accepted for publication in The Astrophysical Journal}

\section{Introduction}

One of the key challenges faced today in cosmology is to understand in
detail how the density distribution of both dark matter and baryons in
the Universe evolved from a relatively smooth initial state at early
times into the non-linear structures we observe today. This non-linear
structure formation is directly coupled to the formation of galaxies
first and later, galaxy clusters. The epoch of reionization (EoR) is a
crucial stage in the history of galaxy and structure formation,
signaling the birth of the first luminous objects as structures first
evolved beyond the well understood linear regime. Although the process
by which the intergalactic medium (IGM) became ionized is quite
complex, the current view is that, when the first proto-galaxies and
quasars form, they ionize the surrounding gas creating the HII
``bubbles''. These regions continue to grow and overlap, so that
eventually all of the neutral gas in the IGM becomes ionized
\citep{barkana01,fan06b}.

Current primary constraints on the epoch of reionization come from two
main sources: the Wilkinson Microwave Anisotropy Probe (WMAP)
determination of the optical depth to recombination through a
late-time signature at large angular scales in the cosmic microwave
background (CMB) polarization spectrum \citep{spergel06,zaldarriaga97a} and
the \Lyman forest absorption spectra towards high redshift quasars
(e.g. \citealt{fan01,fan06a}). This latter Gunn-Peterson effect
\citep{gunn65} is present towards sight lines to quasars out to $z\sim
6.5$ \citep{becker01,cen02} showing that reionization should be ending
by this time (it can also mean that there is a transition in the
Gunn-Peterson optical depth from absorption spectra out to $z\sim 4$
and those out to higher redshifts). However, we note that a small
neutral fraction is enough to completely absorb the \Lyman quasar
flux, so these observations themselves cannot be used to properly
establish the reionization history of the Universe (\citealt{lidz05}).

In terms of the WMAP data, the large angular scale polarization
\citep{page06,dunkley08} yields a Thomson optical depth of
$\tau=0.084\pm 0.016$, so that reionization should happen at $z \sim
10.8 \pm 1.4$ in the favored $\Lambda$CDM cosmological model, if we
assume instantaneous reionization of the Universe
\citep{komatsu08,spergel06}.  The reionization process need not be
instantaneous and if it is less abrupt (see e.g. \citealt{haiman03})
then the Universe may have begun to partly reionize at an even earlier
epoch.  Limited by these two constraints, we know that the
reionization process should have lasted for at least 500 million
years, although there is very little observational evidence on how
this event actually occurred, allowing for various possibilities
including double reionization \citep{cen03,wyithe03a}.

While more precise CMB polarization measurements than with WMAP alone
at large angular scales can provide more information on the
reionization history \citep{holder03, kaplinghat03a,mortonson07}, it is now generally accepted
that detailed information, including the exact history of the
reionization process, will become available with 21-cm signal from the
neutral hydrogen distribution during and prior to reionization
\citep{madau97, loeb04, gnedin04, furlanetto04a, zaldarriaga04,
  sethi05, bharadwaj05, morales03}. Given the line emission, with
frequency selection for observations, the 21-cm data provide a
tomographic view of the reionization process \citep{santos05,
  furlanetto04c} as well as a probe of the dark ages where no luminous
sources are present after recombination at a redshift of 1100
\citep{loeb04}.

We note that small angular scale CMB anisotropies also capture some
information related to reionization, especially the inhomogeneous or
patchy nature of the reionization process
\citep{santos03b,aghanim96,knox98} and through effects such as the
Ostriker-Vishniac \citep{ostriker86,vishniac87} effect.  The 21-cm
background and CMB provide complimentary information related to
reionization since the former is related to the neutral hydrogen
distribution while the latter is due to the electron content \citep{cooray04b}.
Unfortunately, at such small angular scales, the CMB anisotropy
spectrum is rich with a variety of effects contributing to the overall
signal, including galaxy clusters and gravitational
lensing. Therefore, the focus is mainly on 21-cm observations, but
additional information from CMB, such as with a high-resolution version of the EPIC concept mission for the
CMBpol \citep{bock08} may help extract some properties of
the reionization physics.

Motivated by the existing observational constraints and the
possibilities to study reionization through the neutral hydrogen
content with the proposed 21-cm experiments such as the Square
Kilometer Array (SKA\footnote{http://www.skatelescope.org}), the Low
Frequency Array (LOFAR\footnote{http://www.lofar.org}) and the Mileura
Wide-field Array
(MWA\footnote{http://www.haystack.mit.edu/arrays/MWA}), a great deal
of effort has been made recently to understand the 21-cm signal and
its information content (see \citealt{furlanetto06c} for a review).
In parallel with developments in the experimental front, our
theoretical understanding of reionization has also improved both
through numerical simulations and analytical models. Numerical
simulations provide a detailed description of related astrophysics at
these redshifts from first principles by directly solving the
non-linear physics of gravitational collapse, hydrodynamics, and
radiative transfer\citep{gnedin00a,
  razoumov02,sokasian03,ciardi03,kohler05,iliev06,zahn06,mcquinn07}. However,
proper sampling of the epoch of reionization requires simulations with
large volumes $(\sim 100 {\rm Mpc/h})^3$
\citep{barkana04,furlanetto04b} and with adequate mass resolution to
resolve halos where first-light sources are expected to form $(M \sim
10^6 {\rm M_\odot/h})$.  The large volume and the large particle
number makes such simulations computationally expensive, especially in
the context of hydrodynamical calculations related to the gas
physics. The usual approach is to use high resolution, but small
volume simulations or large volume but low mass resolution
simulations. In this paper we make use of a simulation that directly
resolves halos below $10^8\,{\rm M_\odot}$ in a box of 100 Mpc/h,
considered essential to properly take into account the bubble growth
during reionization \citep{shin07}.

Due to the challenging computational requirements of numerical models,
progress on the modeling front has come mostly from analytical
descriptions on the volume filling factor, size distribution of
ionized regions, as well as the power spectrum of the ionized fraction
and density fields
\citep{furlanetto04b,furlanetto06a,sethi05,barkana07}.  These
analytical descriptions have been quite useful to understand the
possible contributions to the 21-cm signal at high redshifts
\citep{barkana05,pritchard06} or under certain simplified conditions
such as the case where spin temperature of neutral gas is
significantly higher than that of the CMB.  The analytical approach is
also crucial to explore the extent to which the full parameter space
of the 21-cm signal and associated cosmology can be established with
data from future surveys planned with MWA, LOFAR, and SKA
\citep{santos06,mcquinn06,mao08}. 

An intermediate approach, based on semi-analytical models combined
with semi-numerical models, has also been developed
\citep{zahn06,mesinger07}. It relies on the generation of realizations
of halo distributions directly from the linear density field and
implementing the corresponding ionization map using criteria similar
to the analytical models. These allow to preserve the spatial
information of the reionization process as provided by simulations,
while achieving a much larger dynamic range than provided by radiative
transfer codes.  

In future, once data become available with the first-generation
low-frequency radio interferometers, it will be useful to have fast techniques to extract the
parameters from the measurements, such as the power spectrum 21-cm brightness temperature.
While detailed numerical simulations to semi-numerical models can be considered for this purpose, it is unlikely such
simulations can be carried out for all variations in parameters of interest, which includes
both astrophysical and cosmological quantities. In this sense, it is more useful to
make use of analytical methods supplemented by well-motivated fitting functions for quantities
such as the power spectrum of ionization fraction during reionization in terms of the power
spectrum of density perturbations that depends on cosmological parameters. This is the approach taken in
predictions related to 21-cm cosmological information content \citep{santos06,mao08}, but
we still need to improve our approximations in such an approach by continuing to study
first-principle numerical models of reionization and 21-cm physics to
test assumptions on the existing analytical models. 

Due to time constraints associated with numerical models,
it is very likely that analytical models are the preferred choice for intensive
astrophysical and cosmological parameter studies from the 21-cm signal
observed with low-frequency radio interferometers. In the case of cosmological parameter estimation with CMB or
from galaxy clustering, the numerical computations provide solutions to an analytical derivation of
either the CMB signal or the dark matter clustering power spectrum. Unfortunately, for the 21-cm anisotropies, such an
approach is likely to be complicated, especially during reionization though at very high redshifts ($\sim 50$), where
physics is simple, a quick numerical calculation of the physics involved can be carried out from first-principles \citep{lewis07}.

In this paper we determine several properties of reionization using a
state of the art large volume and high-resolution simulation of cosmic
reionization based on a photon-advection radiative transfer scheme
combined with a dark matter N-body simulation with recipes for baryons
and star formation \citep{shin07}.  While the simulation itself is
dark matter, the post-processing allows us to convert the
star-formation rate predicted in the simulation to the spin temperature of
the gas in the simulation. Furthermore, we determine the 21-cm
brightness temperature up to $z\sim 25$, by post-processing the
simulation output with a semi-analytical prescription for the X-ray heating of
the gas, the \Lyman coupling, and the collisional coupling, and by taking
fully into account the spatial fluctuations in these quantities. The
existing simulations of the reionization and predictions related to
21-cm signal from such simulations generally ignore the spatial
inhomogeneities associated with X-ray heating and \Lyman radiation
field, though anisotropies of the 21-cm brightness temperature is
expected to be sourced by such inhomogeneities at high redshifts.  We
compare results based on simulations with estimates from a fast
analytical model of reionization
\citep{furlanetto04b,mcquinn05,lidz06}.

Throughout the paper, we make use of the following cosmological
parameters: $\Omega_m=0.26$, $\Omega_\Lambda=0.74$, $\Omega_b=0.044$,
$h=0.72$ and $n_s=0.96$, $\sigma_8=0.77$ based on the results from WMAP, SDSS,
BAO, SN and HST (see \citealt{spergel06} and references therein). The
optical depth is $\tau\approx0.09$, consistent with the WMAP-5 result. The
paper is organized as follows: In the next Section, we outline details
related to the reionization process and compare results from the
simulation and the analytical calculation. We then proceed to describe
how to calculate the corresponding 21-cm signal in Section~3 with
details of the simulation in Section~4. Again we show a comparison of
the results from simulation to analytical models (Section~5). We
conclude with a summary of our results in Section~6.

\section{Cosmic Reionization}

\subsection{Numerical Simulation}
\label{simulation}

In this paper, we make use of one of the largest simulations of cosmic
reionization that has been completed to date \citep{shin07}. We refer
the reader to \cite{shin07} for details related to the hybrid code
that contains a N-body algorithm for dark matter, prescriptions for
baryons and star formation, and a radiative transfer (RT) algorithm
for ionizing photons. We provide a basic summary of the simulation
parameters here as necessary for this study.

The hybrid simulation involves a high resolution N-body calculation 
of $2880^2$ dark matter particles in a $L = 100$ Mpc/h box.
With a particle mass resolution of
$3.02\times10^6\ M_\odot/h$, halos can be reliably resolved down to
masses of $\sim10^8\ M_\odot/h$, accounting for the majority of
photo-ionizing sources.  The simulation distinguishes between the
first generation, Population III (Pop-III) stars and the second
generation, Population II (Pop-II) stars by following the chemical
enrichment of the ISM and IGM as described in \citet{trac06}.  The
input UV spectrum is divided in three energy ranges $13.61\ $eV$<
h\nu\le 24.59\ $eV, $24.59\ $eV$< h\nu \le 54.42\ $eV and $h\nu>
54.42\ $eV, with Pop-II stars with a Salpeter IMF providing 1100, 3830
and 270 ionizing photons per baryon of star formation
respectively. For Pop-III stars with a top-heavy IMF, the corresponding
numbers are 15000, 51500, and 3500 \citep{Schaerer02, Schaerer03}.
The radiative transfer of ionizing photons is calculated
simultaneously as dark matter evolves with the N-body code and with
star formation and baryon physics evolving according to recipes each
step of the way. In this way, our simulation differs from other
descriptions in the literature where radiative transfer and baryon
physics are obtained by post-processing a completed N-body run.

Note that we do not use the halo model of \citet{trac06} for
prescribing baryons and star formation.  Instead, an alternative
approach is taken where we calculate the local matter density $\rho$
and velocity dispersion $\sigma_v$ for each particle.  The baryons are
assumed to trace the dark matter distribution on all scales and we
obtain the local baryon density $\rho_b=\rho(\Omega_b/\Omega_m)$ and
gas temperature $T=\mu\sigma_v2/(3k)$.  Star formation is only allowed
to occur in particles with densities $\rho>100\rho_{\rm crit}(z)$ and
temperatures $T>10^4$ K, thus restricting star formation to regions
within the virial radius of larger halos and these halos are fully
resolved given the low mass resolution of our simulation.

The radiative transfer of ionizing radiation uses a photon-advection
scheme and was run simultaneously with the N-body calculations using a
RT grid with $360^3$ cells.  However, the ionization and recombination
calculations were done for each particle individually rather than on
the grid to preserve small-scale information down to scales of several
comoving Kpc/$h$.  For post-processing, the dark matter, baryons, and
radiation are collected on a grid with $720^3$ cells and the data are
saved every 10 million years from $z=25$ down to $z=5$.

\begin{figure*}[!t]
\centerline{\hspace{1cm}\includegraphics[scale=0.6]{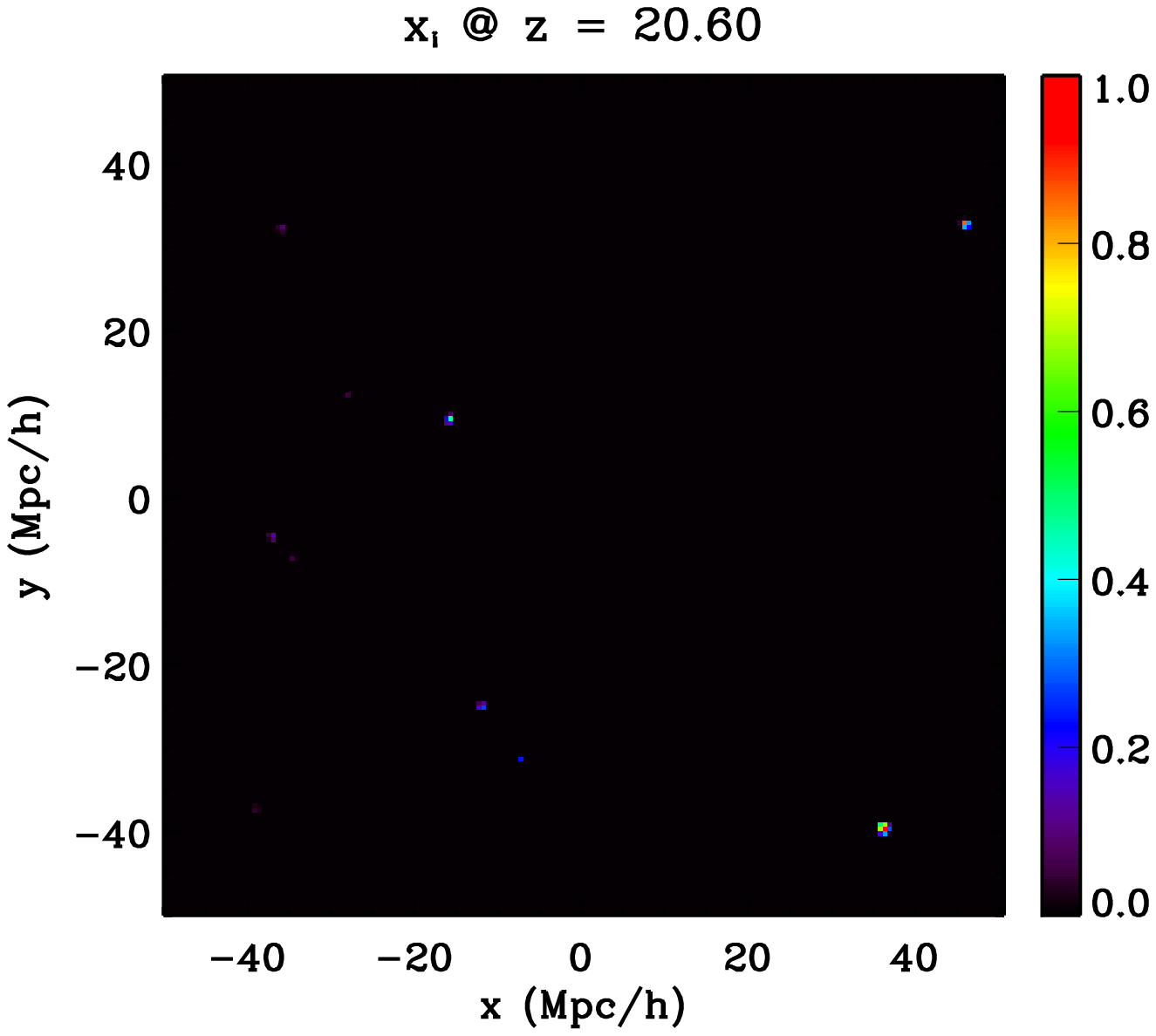}\hspace{-2cm}\includegraphics[scale=0.6]{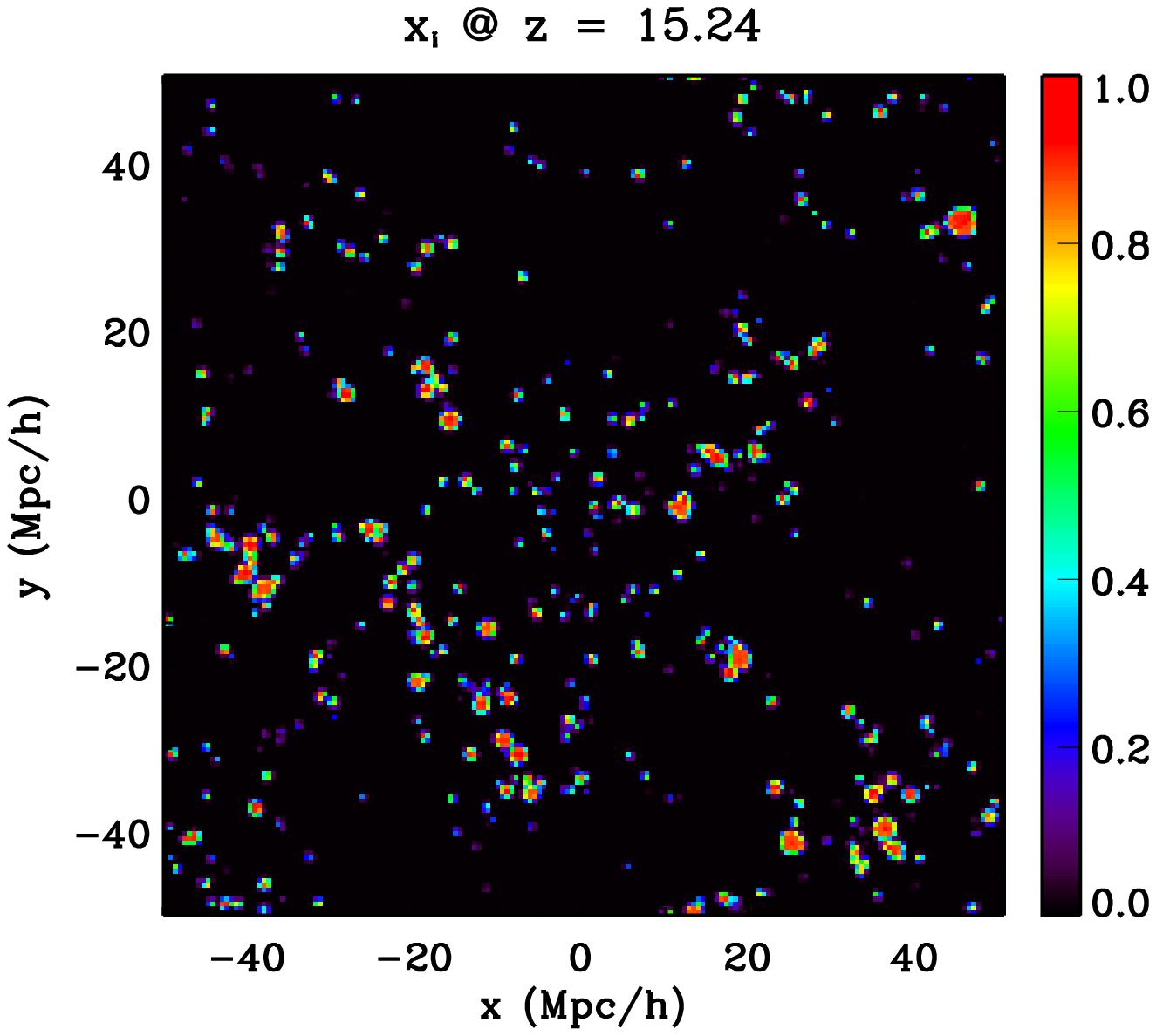}}
\centerline{\hspace{1cm}\includegraphics[scale=0.6]{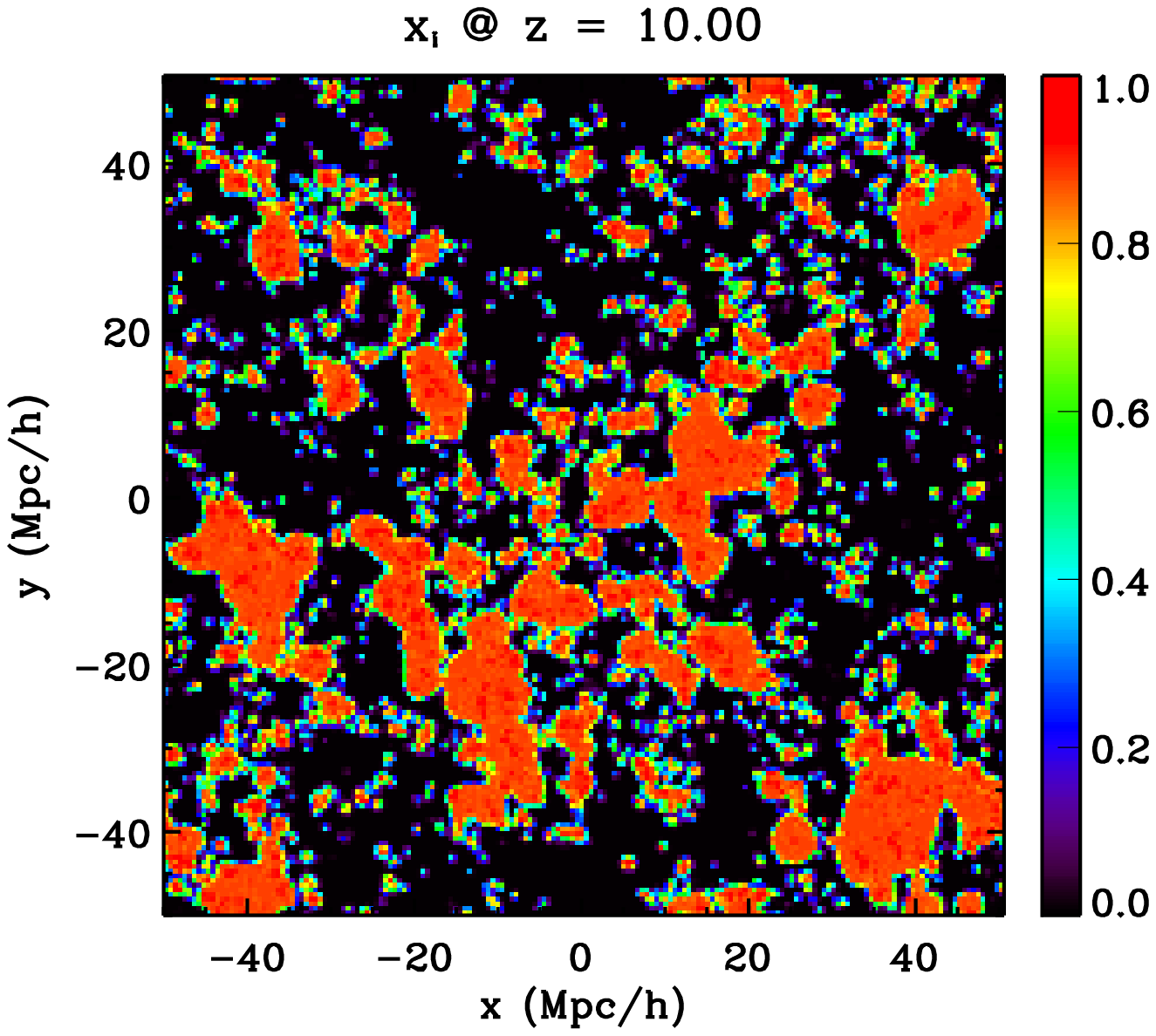}\hspace{-2cm}\includegraphics[scale=0.6]{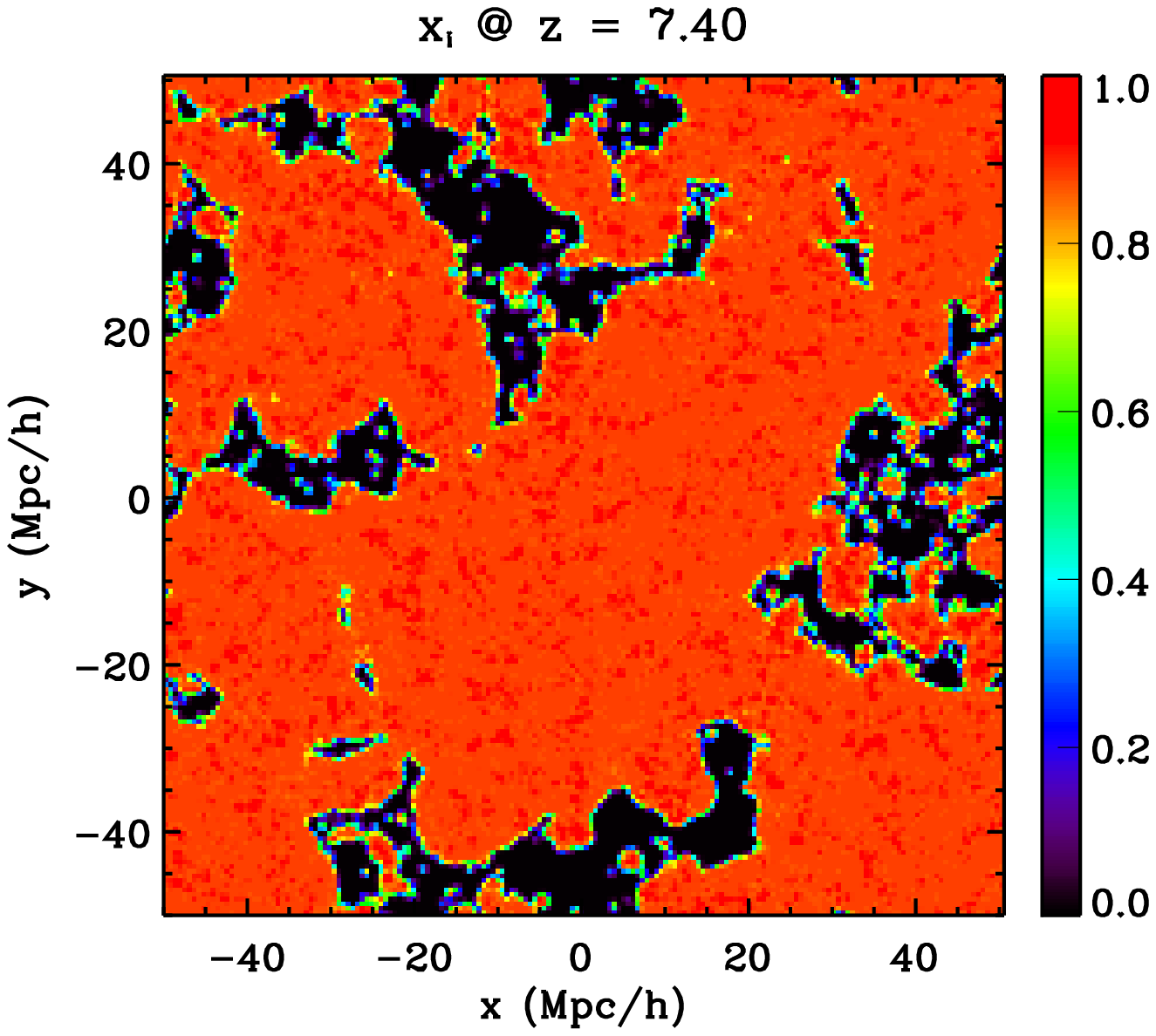}}
\caption{Maps of the ionization fraction from the simulation at redshifts $z=20.60,15.24,10.00,7.40$, corresponding to $\bar{x}_i=0.0002, 0.03, 0.35, 0.84$. Note how there is a clear separation between the highly ionized regions (in red) and the mostly neutral IGM (black).}
\label{map_xi}
\end{figure*}

\begin{figure}[!t]
\hspace{-0.5cm}
\includegraphics[scale=0.75]{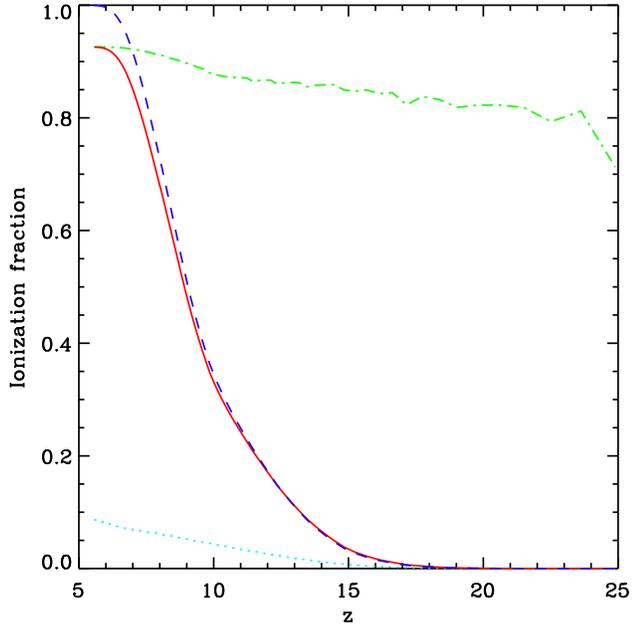}
\caption{The average ionization fraction as a function of redshift 
from the simulation (red solid line) and assuming complete ionization 
of the bubbles (blue dashed
line), defined for $x_i>0.5$. The green dot-dashed line shows the
ionized fraction inside the bubbles while the cyan dotted line shows
the ionized fraction in the IGM (defined has $x_i<0.5$).}
\label{aver_xi}
\end{figure}

We start our analysis of the reionization process by examining a
sequence of cuts through the simulation box of the fraction of free
electrons $x_i$ (the ionization fraction), shown in Figure
\ref{map_xi}. 
In this simulation box, reionization begins roughly at a redshift of
$\sim 18$.  The ionized bubbles have complex shapes and cannot be
simply described with spherical models for the 3-dimensional HII
regions surrounding UV sources. Complete overlap of the ionization
patches occurs at the redshift of $z \sim 6$. 
Note that although the ionization fraction can have a range of values between 0 and 1, most
of the volume in the simulation is almost completely ionized (very
high $x_i$) or completely neutral, thus favoring the current view of
reionization based on the percolation of large ionized bubbles. Figure
\ref{aver_xi} shows the evolution of the average ionization fraction
compared with the values if we assume that gas is completely ionized
inside bubbles ($x_i=1$) and completely neutral outside ($x_i=0$).
Bubbles are defined by the threshold $x_i>0.5$.

\subsection{Analytical models}
\label{sec:models}

Our analytic model for reionization, which we compare with results
from our numerical simulation, follows the approach of
\citet{furlanetto04b}.  The mass of the ionized gas is linked to the
mass in galaxies by the ansatz, $m_{\rm ion}=\zeta m_{\rm gal}$, where
$\zeta$ is an ionizing efficiency. A spherical region of gas of mass
$m$ is considered ionized if it contains sufficient sources to self
ionize, i.e. $f_{\rm coll}\ge\zeta^{-1}$. In the excursion set
formalism this criteria is well described by a mass dependent linear
barrier $B(m,z)=B_0+B_1 \sigma^2(m)$, where $\sigma^2(m)$ is the
variance of the density fluctuations on the scale $m$.  With this we
can calculate the mass function of bubbles (the comoving number
density of HII regions with masses in the range $m\pm dm/2$):
\begin{equation}
  \frac{\ud n(m)}{\ud m}dm=\sqrt{\frac{2}{\pi}}\frac{\bar{\rho}}{m^2}\left|\frac{\ud\log\sigma}{\ud\log m}\right|\frac{B_0}{\sigma(m)}\exp\left[-\frac{B^2(m,z)}{2\sigma^2(m)}\right]dm\, ,
\end{equation}
where $\bar{\rho}$ is the mean mass density of the Universe.

Note that we renormalize the resulting mass function to enforce the
requirement $\bar{Q}=\zeta f_{\rm coll}$, where $\bar{Q}$ is the
filling fraction of bubbles. Next we must determine the appropriate
value for the ionization efficiency $\zeta$.  We allow $\zeta$ to vary
with redshift and require that $x_i=\zeta f_{\rm coll}$ at all
redshifts.  Here $x_i$ is determined from the simulations so that we
bypass the need for a source prescription when defining $\zeta$ and we
assume a Press-Schechter mass function when determining $f_{\rm
  coll}$.  In principle, using the Sheth-Tormen mass function gives a
weaker redshift dependence for $\zeta$, but we use Press-Schechter for
greater consistency with the reionization model of
\citet{furlanetto04b}.

To calculate fluctuations in the 21-cm brightness temperature, we
first need to calculate the correlation functions in the ionization
fraction $\xi_{x_i x_i}$, density $\xi_{\delta\delta}$, and the
cross-correlation of these two quantities $\xi_{x_i\delta}$, where \be
\xi_{ab}\equiv \left<(a-\bar{a})(b-\bar{b})\right>, \ee and
$\delta=\rho/\bar{\rho}-1$.  We use the halo model to calculate
$\xi_{\delta\delta}$ \citep{cooray02}. \citet{furlanetto04b} present
an ad hoc model for the correlation functions $\xi_{xx}$ and
$\xi_{x\delta}$, designed to ensure that the correct limiting behavior
as $x_H\rightarrow 0,1$ is obeyed.  A fundamental problem with their
approach is that bubbles are assumed to be spherical at all times
leading to problems describing the overlap of bubbles properly.
\citet{mcquinn05} later attempted to modify the \citet{furlanetto04b}
model to forbid bubble overlap.  However, since neither of these
models correctly handles bubble overlap, we choose to use the original
formulation of \citet{furlanetto04b}.  However, we incorporate the
corrected calculation of the bubble bias, as noted by
\citet{mcquinn05}.  We note that a more physically motivated method
based upon the two-step approximation has recently been developed by
\citet{barkana07}, but for purposes of the present discussion where we
are investigating the extent to which a simple model for parameter
estimates can be compared to numerical simulations and is found
in general to provide good agreement, such details are
unnecessary given the availability of simulations.

\begin{figure*}[!t]
\centerline{\includegraphics[scale=0.55]{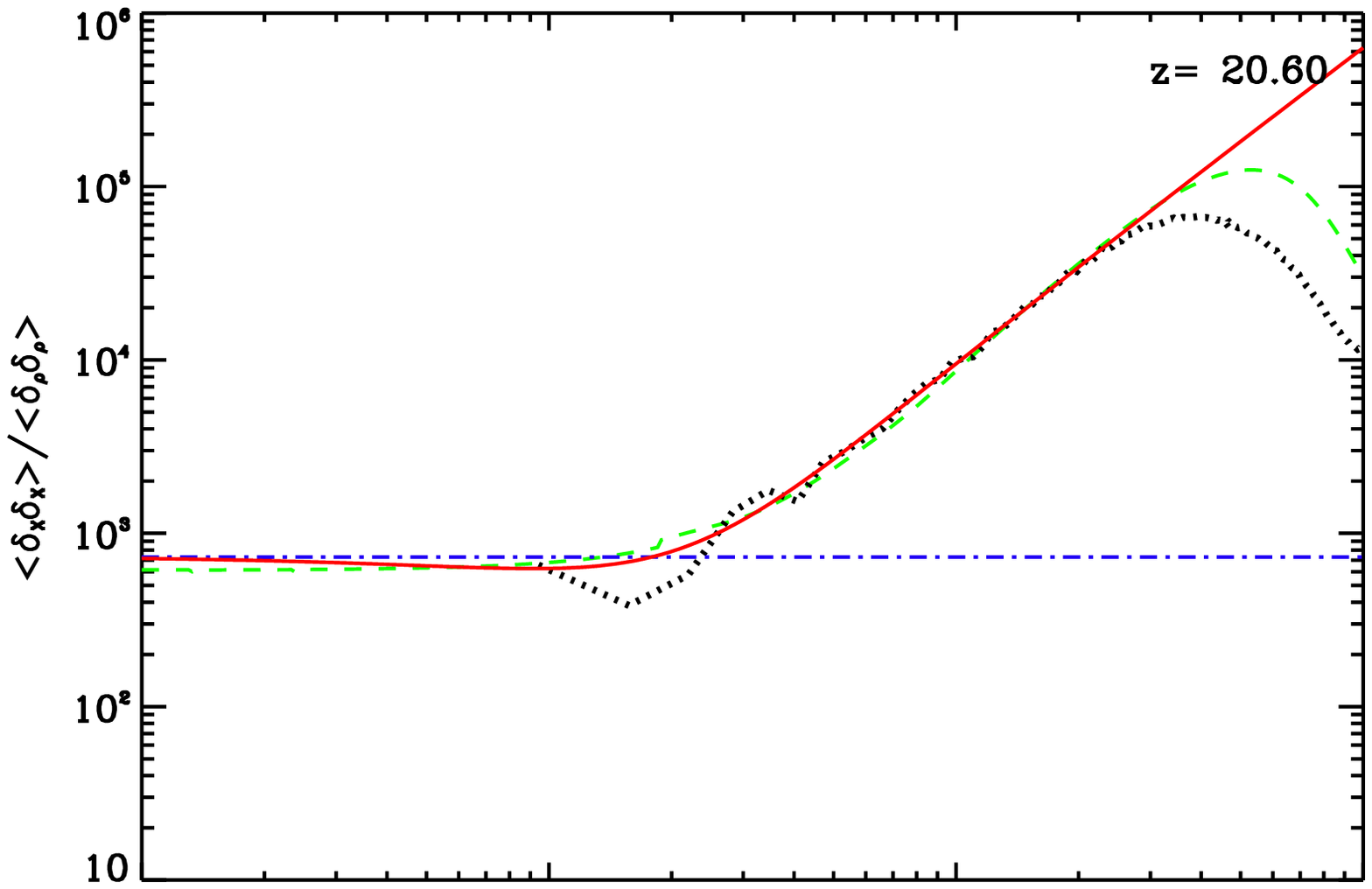}\hspace{-2cm}
  \includegraphics[scale=0.55]{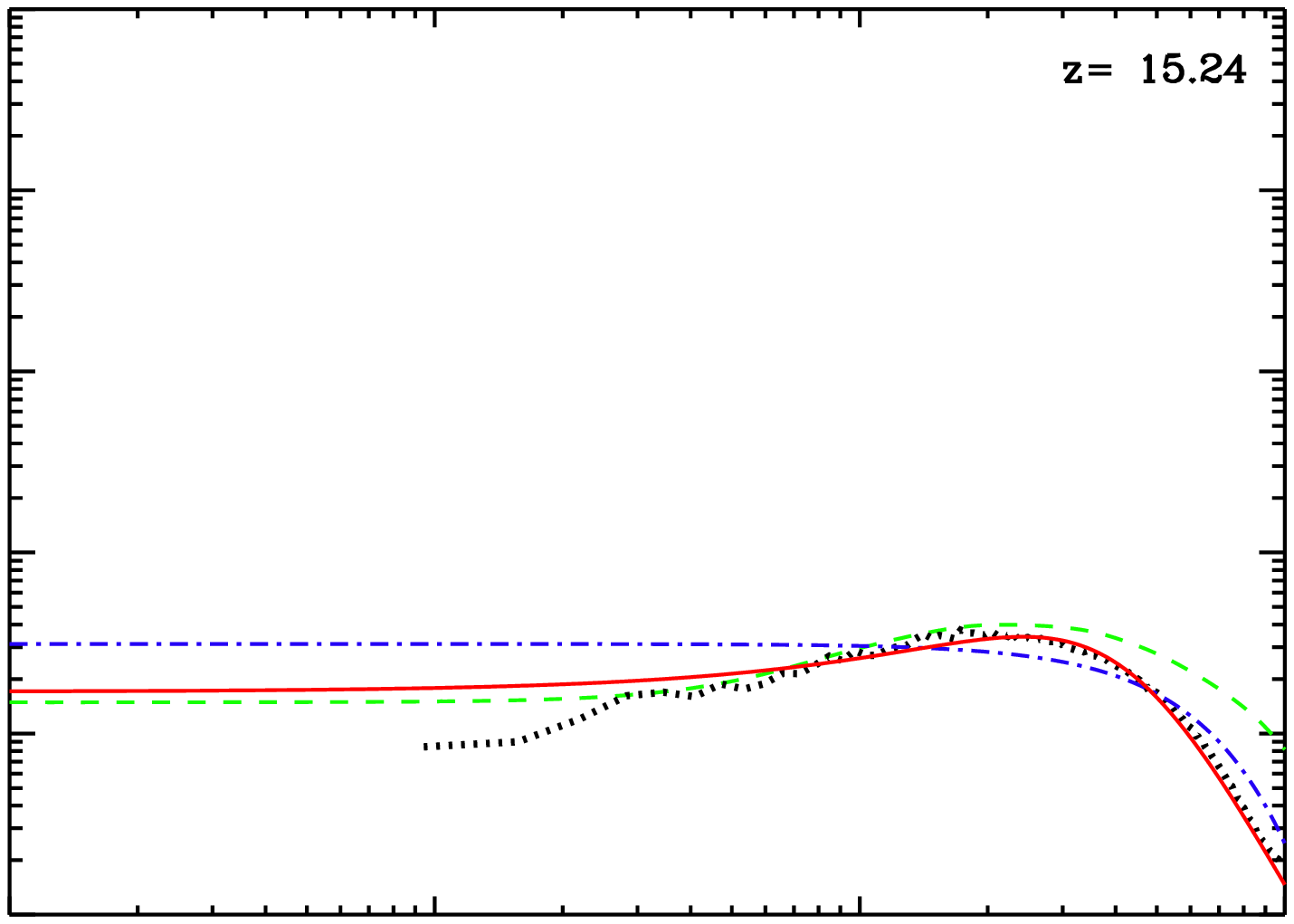}}
\vspace{-1.2cm}
\centerline{\includegraphics[scale=0.55]{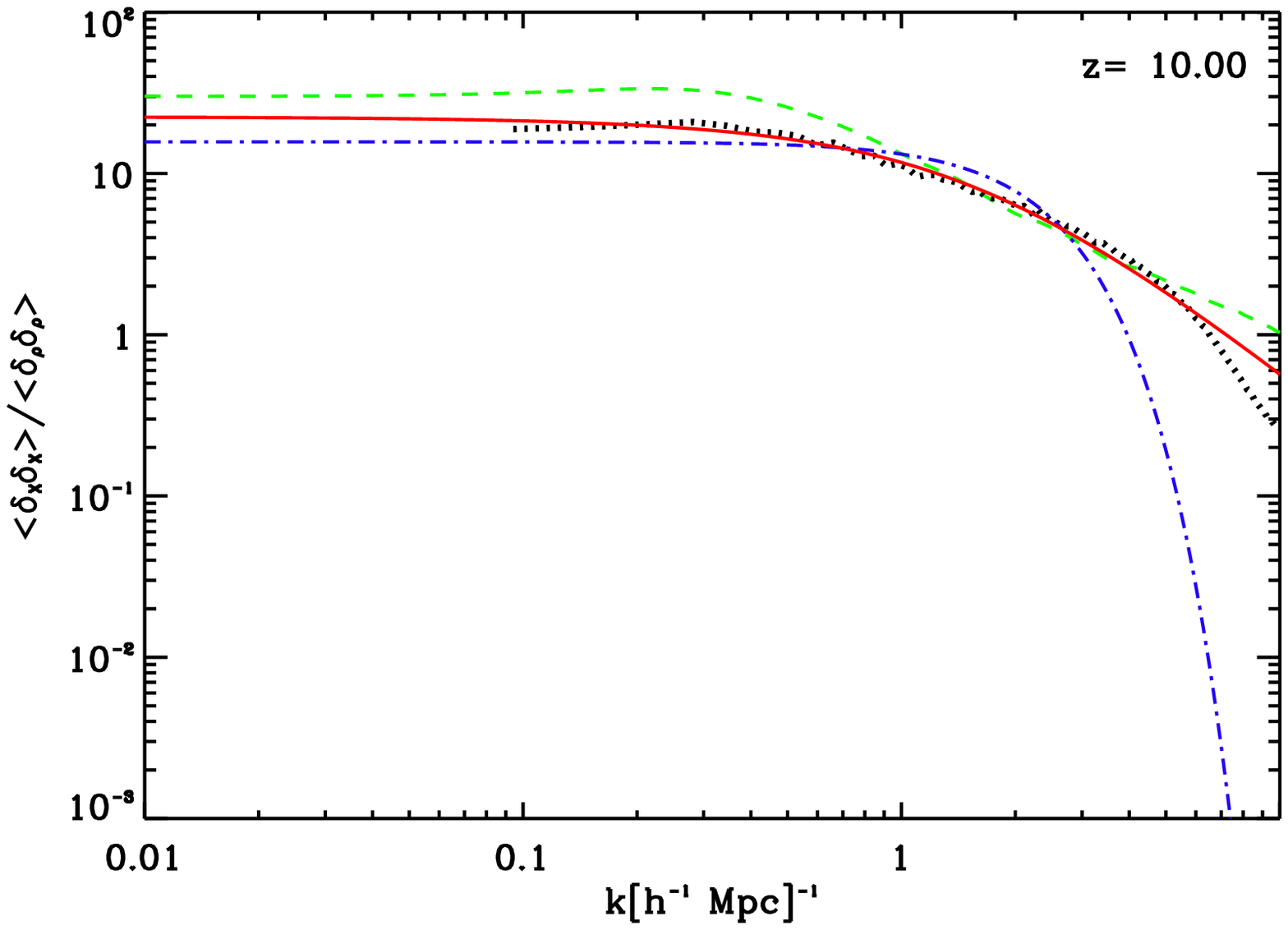}\hspace{-2cm}
  \includegraphics[scale=0.55]{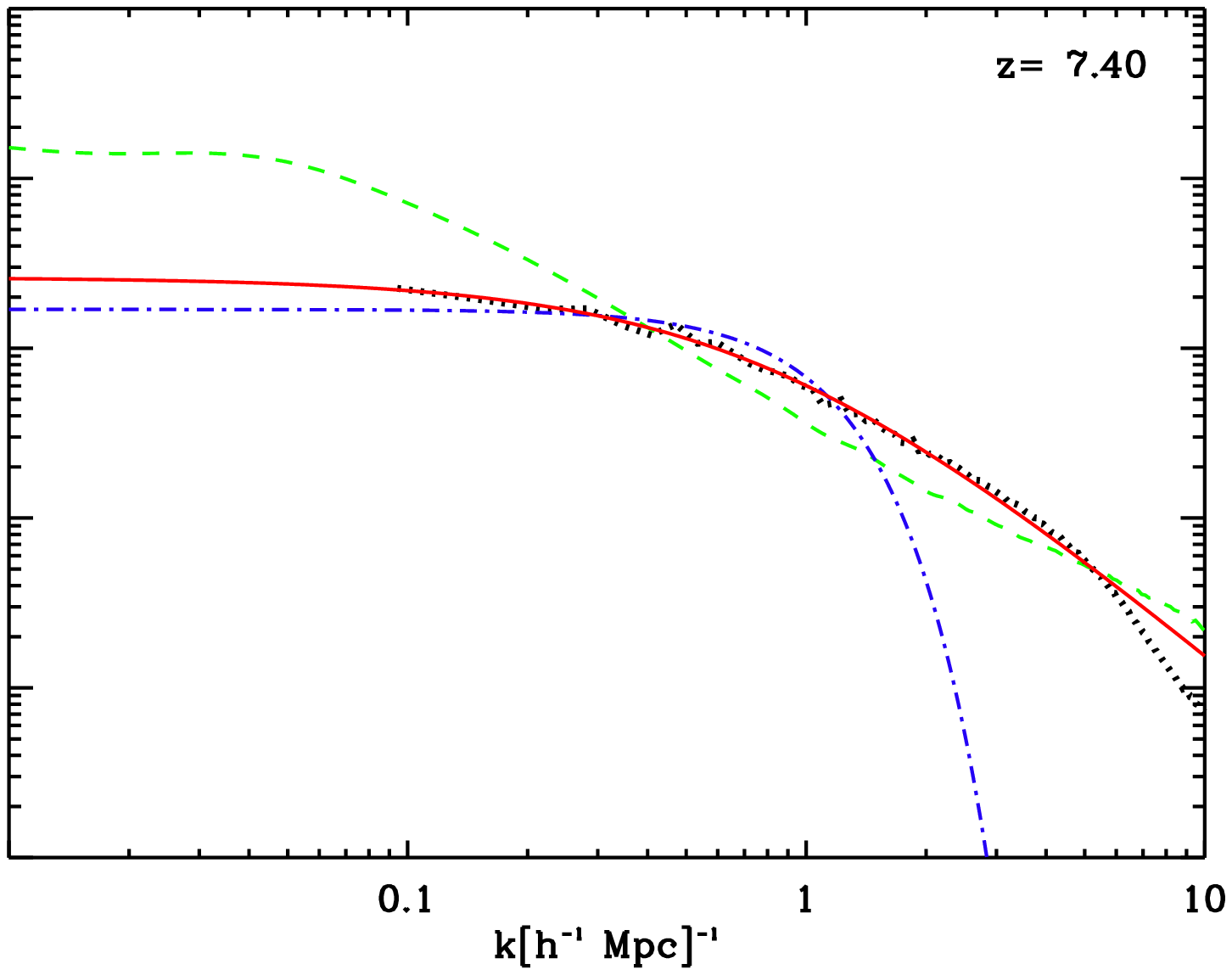}}
\caption{Ratio of the power spectra of the ionization fraction,
  $\delta_{x_i}\equiv x_i/\bar{x_i}-1$ to the matter one,
  $\delta\equiv \rho/\bar{\rho}-1$. The simulation corresponds to the
  dotted (black) line while the dashed-green line gives the results of
  the analytical model followed in the paper (based on
  \citealt{furlanetto04b}). Dot-dashed (blue) lines are fitting curves
  using the two parameter model suggested in \citet{santos06} while
  solid (red) curves use the model in \citet{mao08}. Redshifts, as
  labeled from top-left to bottom-right are: $z=20.6, 15.2, 10.0,
  7.4$, corresponding to $\bar{x}_i=0.0002, 0.03, 0.35, 0.84$
  respectively.
}
\vspace{0.5cm}
\label{reion_Pk}
\end{figure*}

The 3-dimensional power spectra of our simulation were performed using
the fast Fourier transform package fftw-3.1.1\footnote{
  http://www.fftw.org} and is defined through: $<a(\bar{k})
  b^*(\bar{k}')>\equiv (2\pi)^3 \delta^3(\bar{k}-\bar{k}')P_{ab}(k)$. 
We then binned our modes with $\delta k = {2\pi \over 100}
({\rm Mpc}/h)^{-1}$ and computed the average power spectra in each
bin.  Throughout the paper we will plot the dimensionless power
spectrum, $\Delta^2=k^3 P(k)/2\pi^2$, which gives the contribution to
the variance per logarithmic interval in k.  In order to test the
analytical calculation, we show in Figure~\ref{reion_Pk} the ratio of
the 3-dimensional power spectrum of the ionization fraction to the
matter density, for the simulation and analytical models.

The features in the power spectrum of ionization fraction relative to
density perturbations can be described as following. At scales much
larger than the bubble size, the ionization fraction power spectrum
is proportional to the matter density with an overall scaling which
can be assigned to the bias factor of the ionized regions.  From the
simulation results shown in dotted lines this can be seen more easily
at high redshifts where bubble sizes are small.  At these redshifts
($z=20.60$ and $z=15.24$), there is an increase in the ionization
power spectrum at small scales due to the Poisson behavior of the
bubble distribution.  This is analogous to the shot-noise component of
the galaxy power spectrum even at low redshifts that dominate
fluctuations at non-linear scales.

As we move to lower redshifts and larger bubbles,
Figure~\ref{reion_Pk} shows more clearly that the power spectrum of
the ionization fraction $P_{x_ix_i}$ decreases on physical scales
smaller than the typical bubble size, due to the smoothing effect of
bubbles. Note however that the decrease is not abrupt since there is a
distribution of bubble sizes at any redshift with varying ionization
fractions. The analytical calculation we discussed so far (dashed
line) seems to agree reasonably well with the simulation, although at
larger ionization fractions ($x_i>0.8$), when the bubbles occupy most
of the simulation volume, there are some differences.  This difference is probably due
to the finite size of the simulation box which effectively limits the bubble
size and reduces power on large scales. Note however that even if we our simulations involved larger volumes,
the power could continue to be smaller than in the analytical case since the simulation accounts for 
self-shielding of dense regions that remain neutral. This causes the bubble growth to stall and in return limit the maximum
bubble size \citep{furlanetto05}.

In Figure~\ref{reion_Pk}, in addition to the comparison between
results from our numerical simulation and a widely-used analytical
model, we also compare with two simple fitting functions of the
ionization fraction given the density field power spectrum. We do this
comparison since when exploring the full parameter space probed by
21-cm experiments \citep{santos06,mao08}, calculations make use of
fitting formulae to describe the power spectrum of the ionization
fraction without making use of detailed analytical models or numerical
calculations of bubble growth. Here, we compare our results to two
fitting functions in the literature where the power spectrum of the
ionized fraction $P_{x_ix_i}$ is given as
\begin{eqnarray}
P_{x_ix_i}(k) &\ =\ & b_{x_ix_i}^2e^{-(kR_{x_ix_i})^2}P_{\delta \delta}(k) \label{eqn:pxx1}\\ 
P_{x_ix_i}(k) &\ =\ & b_{x_ix_i}^2\left[1+\alpha_{x_ix_i}(kR_{x_ix_i})+\right.\\ \nonumber
&&\left. +(kR_{x_ix_i})^2\right]^{-\gamma/2}P_{\delta \delta}(k)
\label{eqn:pxx2}
\end{eqnarray}
where the first approximation is from \citet{santos06} and the second
is from \citet{mao08} and $b_{x_ix_i}$, $\alpha_{x_ix_i}$,
$\gamma$, and $R_{x_ix_i}$ are free parameters that are varied to obtain a fit
to the simulation results. In Table~1, we list the values that were
obtained by comparing numerical simulations with the above form.

\begin{table}[h!]
\begin{center}
\begin{tabular}{lccccc}
\hline
Function & Parameter & z=20.6 & z=15.2 & z=10.0 & z=7.4\\
\hline \hline 
Eq.~\ref{eqn:pxx1} & $R_{x_ix_i}$  (h$^{-1}$Mpc) & 0  & 0.16 & 0.42 & 0.96\\
                  & $b_{x_ix_i}$           & 27    & 17.7   & 4.0 & 1.3\\
\hline
Eq.~\ref{eqn:pxx2} & $R_{x_ix_i}$ (h$^{-1}$Mpc) & -4.53 & 0.23 & 0.62 & 1.24\\
 &              $\alpha_{x_ix_i}$  & 0.83  & -1.12 & 0.93 & 1.48 \\
  & $\gamma$  & -1.77  & 3.72 & 1.93 & 1.99 \\
                  & $b_{x_ix_i}$           & 27.2    & 13.1   & 4.7 & 1.6\\
\hline
\end{tabular}
\caption{Parameters of the ionization fraction power spectrum
  model described in equations \ref{eqn:pxx1} \& \ref{eqn:pxx2} fitted
  to our simulation power spectrum measured at $z=20.6,15.2,10$, and 7.4.
Note that to get the ``physical'' $R_{x_ix_i}$ based on our definitions one should multiply 
the above values by a factor of $(2\pi)$.}
\label{tab:rval}
\end{center}
\end{table}
While these two fitting functions have been used in the literature
when making predictions related to how well cosmological parameters
can be measured with 21-cm interferometers such as MWA and LOFAR, as
can be seen from Figure~\ref{reion_Pk}, both these functions are not
completely accurate descriptions of the power spectrum of ionized
fraction over all the redshift range we have studied with simulations.
This is due to the fact that, for example, the first approximation
from \citet{santos06} assumes that bubbles can be described with a
single size leading to a sharp cut-off at the wavenumber corresponding
to the inverse radius, while from simulations and in the analytical
model of \citet{furlanetto04b} the bubbles have varying sizes.  At $z
< 15$, when bubbles have started to grow, the fitting formula of
\citet{mao08} provides an improved fit given the additional freedom
provided by the parameters $\alpha$ and $\gamma$ and this description
is probably adequate enough for now when making predictions related to
the extent to which cosmological parameters can be measured with a
21-cm experiment. On the other hand, when model fitting real data, it
may be necessary to improve the estimate of $P_{x_ix_i}(k)$ beyond
simple fitting functions as the ones listed in
equations~(\ref{eqn:pxx1}) and (\ref{eqn:pxx2}).  In this respect, we
note that the analytical model of \citet{furlanetto04b} provides a
more accurate description of the results from our numerical
simulation and a more clear interpretation of the values obtained. 
If such a model can be further improved to quickly explore
a large parameter space in a reasonable time, it may be useful to implement 
such a model in a numerical code for parameter estimates, such as based on the
Markov Chain Monte-Carlo technique, instead of simply using fitting functions to estimate parameter
values from 21-cm interferometers.

\begin{figure*}[!t]
\centerline{\hspace{-0.5cm}\includegraphics[scale=0.7]{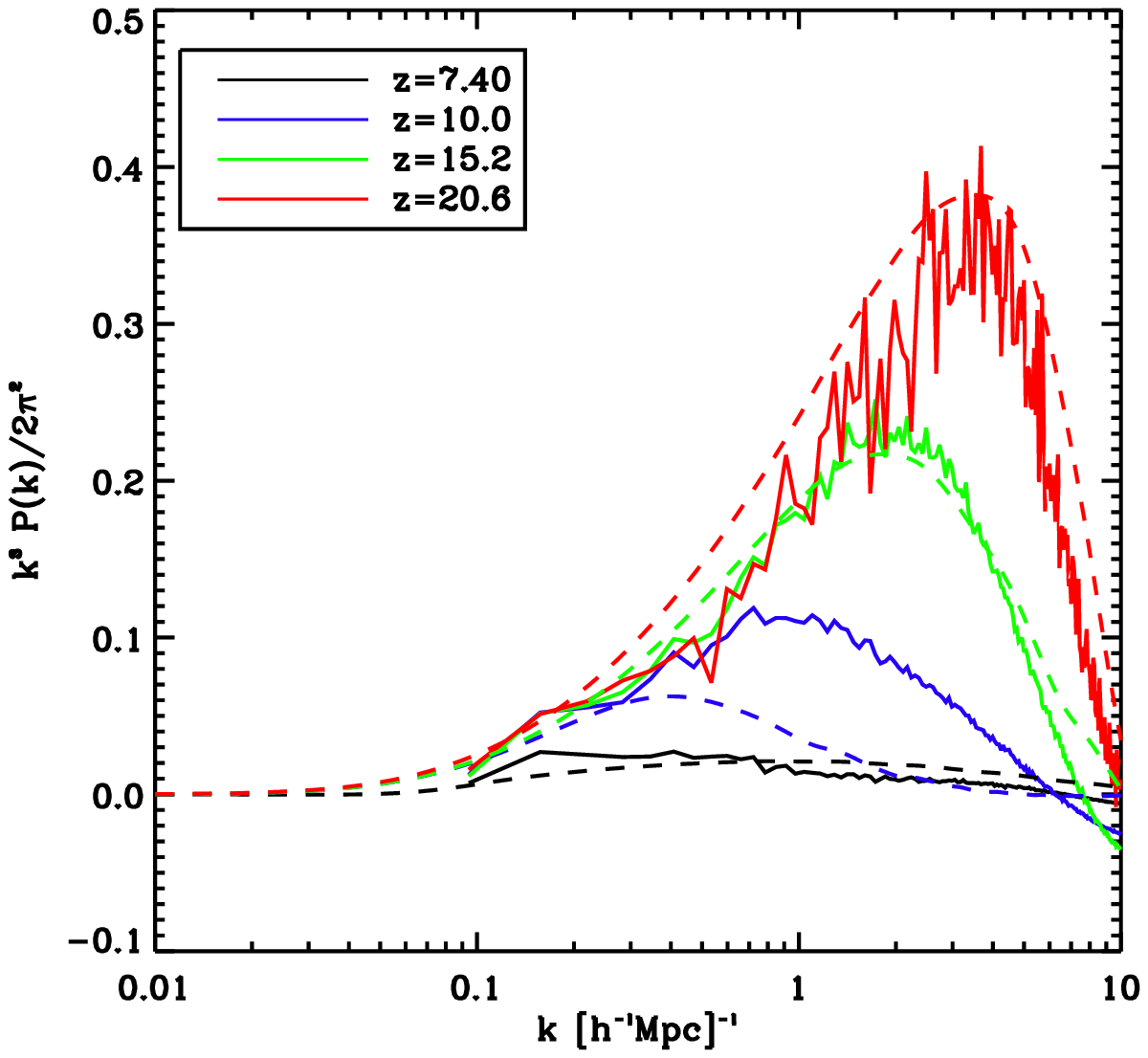}\hspace{0.3cm}  \includegraphics[scale=0.7]{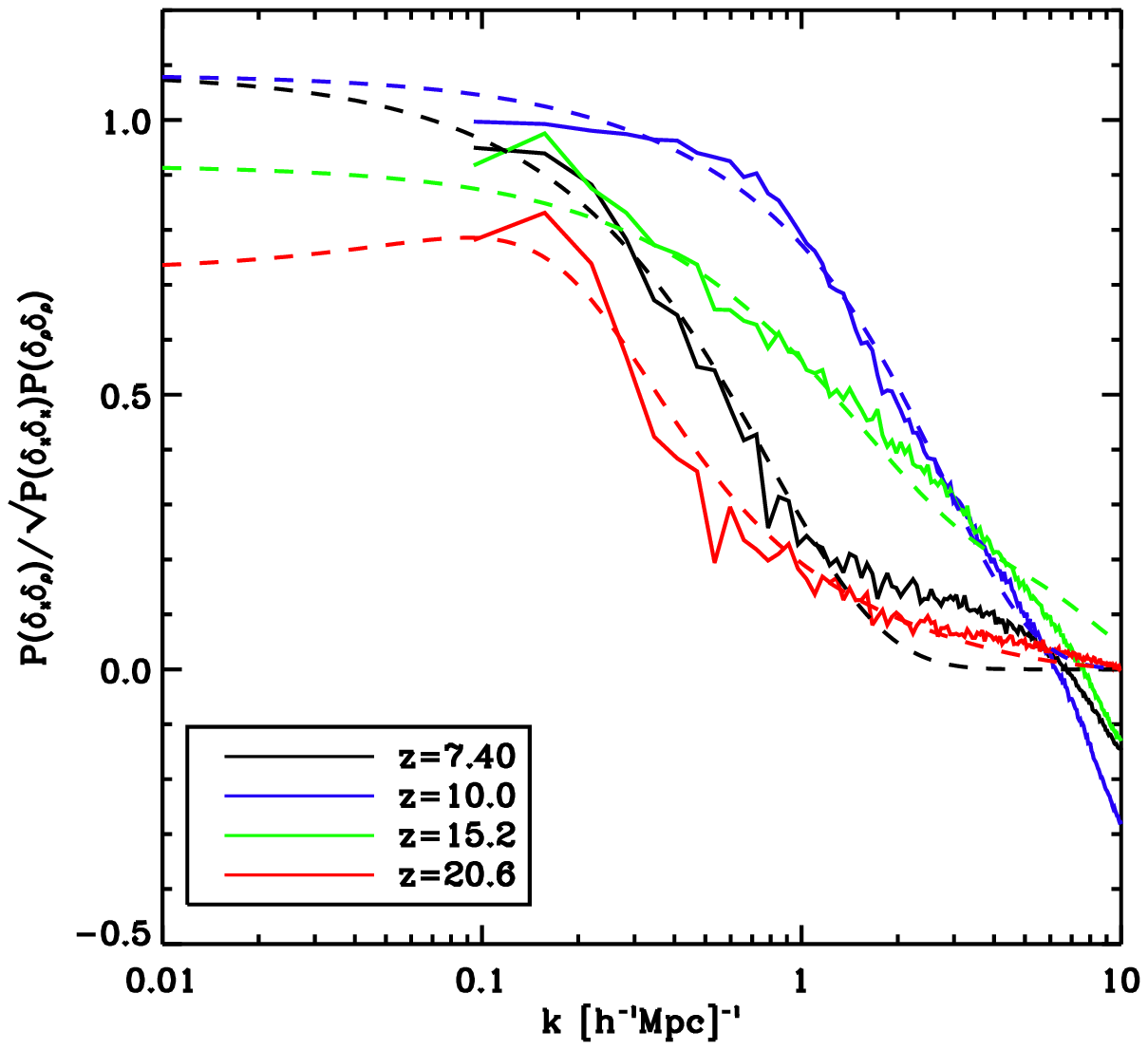}}
\caption{Left: Cross
    power spectra of  the ionization fraction and matter density
  ($\left<\delta_{x_i}\delta\right>$) from simulations. 
Solid lines give the simulation
  values while dashed lines are for the analytical model from
    Furlanetto et al. From top to bottom the curves correspond to
  $z=20.6, 15.2, 10.0, 7.4$.  Right: The correlation
  coefficient, $r=P_{x_i\delta}/\sqrt{P_{x_ix_i}P_{\delta\delta}}$, with
  simulations (solid line) and the fitting function of \citet{mao08} (dashed line)
for the same redshifts as the left panel.}
\vspace{0.5cm}
\label{xr_corr}
\end{figure*}

Figure~\ref{xr_corr}, left panel, shows the cross power spectrum between
the ionization fraction and the density field.  The peak of this
cross-power spectrum moves to larger scales as the redshift decreases
since it is related to the typical size of bubbles during
reionization.  At smaller scales, there is some indication that the
cross-correlation power spectrum becomes negative in the numerical
simulation suggesting more of an anti-correlation than what is seen in
the analytical model.  We believe this is partly due to small neutral
regions inside HII bubbles because the simulation takes into account
the self-shielding of dense regions. Note
however, that as reionization progresses and the radiation intensity
become larger, the existence of small fingers from large bubbles
protruding into the now small neutral regions will play an important
part in this anti-correlation.

Similar to equation~(\ref{eqn:pxx2}), we can also write the cross power
spectrum between $x_i$ and $\delta$ assuming a perfect correlation of
the two fields such that
\begin{eqnarray}
  P_{x_i\delta} &=& \sqrt{P_{x_ix_i}P_{\delta \delta}} \, ,
\label{eqn:pxdelta}
\end{eqnarray}
and by making use of the fitting formulae for the ionization fraction
power spectrum from equation~(\ref{eqn:pxx1}), as was used by
\citet{santos06}. In \citet{mao08}, this cross power spectrum is
modeled as $P_{x_i\delta} =b^2_{x_i\delta}
\exp[-\alpha_{x_i\delta}(kR_{x_i\delta})-(kR_{x_i\delta})^2]P_{\delta
  \delta}$, and in Figure~\ref{xr_corr} right panel we also show this
case with a new set of parameters
$(b_{x_i\delta},\alpha_{x_i\delta},R_{x_i\delta})$.  Here, we
plot the correlation coefficient $r=P_{x_i\delta}
  /\sqrt{P_{x_ix_i}P_{\delta \delta}}$ and we compare the fitting
function motivated by \citet{mao08} to numerical simulations.
While there is a good overall agreement between the
simulations and the analytical model, we find differences at
$z \sim 20$ between the two descriptions. Such a disagreement,
however, is not a concern for first-generation 21-cm observations
since these instruments mostly concentrate on $z\sim 6$ to $9$ 
during reionization.

\section{The 21-cm signal: Theory}

We present here the overall calculation of the 21-cm signal. Details
specific to the simulation and the analytical calculation considered in this
paper will be given later in the appropriate sections.

\subsection{Brightness temperature}

One of the best ways to observe the reionization process is through
the 21-cm brightness temperature, corresponding to the change in the
intensity of the CMB radiation due to absorption or emission when it
travels through a patch of neutral hydrogen.  It is given, at an
observed frequency $\nu$ in the direction $\bn$, by \bea \delta
T_b(\bn,\nu) & \approx & \frac{T_S - \tcmb}{1+z} \, \tau
\label{eq:dtb} 
\eea where $T_S$ is the temperature of the source (the spin
temperature of the IGM), $z$ is the redshift corresponding to the
frequency of observation ($1+z=\nu_{21}/\nu$, with $\nu_{21} = 1420$
MHz) and $\tcmb = 2.73 (1+z) K$ is the CMB temperature at redshift
$z$.  The optical depth, $\tau$, of this patch in the hyperfine
transition \citep{field59} is given in the limit of $k_B T_s >> h
\nu_{21}$ by \bea \tau & = & \frac{ 3 c^3 \hbar A_{10} \, n_{\rm
    HI}}{16 k \nu_{21}^2 \, T_S \, (1+z)(\partial V_r/\partial r) },
\eea where $A_{10}$ is the spontaneous emission coefficient for the
transition ($2.85 \times 10^{-15}$ s$^{-1}$), $n_{\rm HI}$ is the
neutral hydrogen number density and $\partial V_r/\partial r$ is the
gradient of the total radial velocity along the line of sight (with
$V_r\equiv \bV\cdot\bn$); on average $\partial V_r/\partial r =
H(z)/(1+z)$. In this paper we will neglect perturbations from the
peculiar velocity of the gas. The neutral density can be expressed as
$n_{\rm HI}=f_{\rm HI}X\rho_b/m_p$ where $f_{HI}=\rho_{\rm
  HI}/\rho_{\rm H}$ is the fraction of neutral hydrogen (mass
weighted), $X\approx 0.76$ is the hydrogen mass fraction, $\rho_b$ is
the baryon density and $m_p$ the proton mass.  The 21-cm temperature is
then: \bea
\label{t21}
\delta T_b(\bn,\nu) & \approx & 23 f_{\rm HI}\frac{\rho_b}{\bar\rho_b}
\left(1-{\tcmb\over T_S}\right)
\left( \frac{h}{0.7}\right)^{-1}\times\nonumber\\
&&\left( \frac{\Omega_b h^2}{0.02} \right) \left[
  \left(\frac{0.15}{\Omega_m h^2} \right)\left(\frac{1+z}{10}\right)
\right]^{1/2}\,\rm{mK} \eea In order to proceed, we will need a
prescription to calculate the spin temperature of the gas.

\subsection{Spin temperature}

The spin temperature is coupled to the hydrogen gas temperature
($T_K$) through the spin-flip transition, which can be excited by
collisions or by the absorption of \Lyman photons (Wouthusysen-Field
effect) and we can write: \be 1-{\tcmb\over T_S}={y_{tot}\over
  1+y_{tot}}\left(1-{\tcmb\over T_K}\right), \ee where
$y_{tot}=y_\alpha+y_c$ is the sum of the radiative and collisional
coupling parameters and we are already assuming that the color
temperature of the \Lyman radiation field at the \Lyman frequency is
equal to $T_K$.  When the coupling to the gas temperature is
negligible (e.g. $y_{tot}\sim 0$), $T_S\sim \tcmb$ and there is no
signal. On the other hand, for large $y_{tot}$, $T_S$ simply follows
$T_K$.

Collisions can be important for decoupling the HI 21-cm spin temperature from the CMB,
especially at high redshifts \citet{nusser05} and the coupling coefficient is given by
\begin{equation}
  y_c=\frac{4T_\star}{3A_{10}T_\gamma}\left[\kappa^{HH}_{1-0}(T_k)n_H +\kappa^{eH}_{1-0}(T_k)n_e\right],
\label{yc}
\end{equation}
where $T_\star\equiv hc/k\lambda_{21 \rm{cm}}=0.0628\,\rm{K}$,
$\kappa^{HH}_{1-0}$ is tabulated as a function of $T_k$
\citep{allison69,zygelman05}, $\kappa^{eH}_{1-0}$ is taken from
\citet{furlanetto07} and $n_e$ is the electron number density (see
also \citealt{kuhlen06}). For a more detailed analysis of the
collisional coupling, see \citet{hirata07}.

The Wouthysen-Field effect \citep{wouthuysen52,field59} coupling is given by
\begin{equation}\label{yalpha}
  y_\alpha=\frac{S_\alpha J_\alpha}{J_c},
\end{equation}
with
\begin{eqnarray}
J_c &\equiv&  \frac{16\pi^2T_\star e^2 f_\alpha}{27A_{10}T_\gamma m_e c} \\
&\approx& 5.552 \times 10^{-8}(1+z) \quad \quad {\rm m^{-2} s^{-1} Hz^{-1} sr^{-1}} \nonumber
\end{eqnarray}
where $f_\alpha=0.4162$ is the oscillator strength of the \Lyman
transition and $(1+z)$ comes from $T_\gamma$. 
In equation~\ref{yalpha}, $S_\alpha$ is a correction factor of order unity, which
describes the detailed structure of the photon distribution in the
neighborhood of the \Lyman resonance \citep{chen04,hirata06,
  chuzhoy07,furlanetto06d}.  We make use of the approximation for
$S_\alpha$ outlined in \citet{furlanetto06c}.  The proper \Lyman
photon intensity, $J_\alpha$ (the spherical average of the number of
photons hitting a gas element per unit proper area per unit time per
unit frequency per steradian) is given by a sum over the hydrogen
levels $n$, 
\bea
J_{\alpha}(\bx,z) \ &=&\ \frac{(1+z)^2} {4 \pi}
\sum_{n=2}^{n_{\rm max}} f_{\rm rec}(n)\times\\ \nonumber
&\times& \int {d\Omega'\over 4
  \pi}\int_0^{x_{\rm max}(n)} dx'\ \epsilon_\alpha(\bx+\bx',\nu_n',
z')\ ,
\label{Jax} 
\eea
where $f_{\rm rec}(n)$ is the fraction of Lyman-n photon that
cascade through \Lyman and $\epsilon(\bx,\nu_n, z)$ is the comoving
photon emissivity, defined as the number of photons emitted at
position $\bx$, redshift $z$ and frequency $\nu$ per comoving volume,
per proper time and frequency. Note that the redshift $z'$ in equation
\ref{Jax} is such that $x'=\int_z^{z'} cH^{-1} dz''$.  The absorption
at level $n$ at redshift $z$ corresponds to an emitted frequency at
$z'$ of \beq \nu_n' = \nu_{\rm LL} (1-n^{-2}) {(1+z')\over (1+z)}\ ,
\label{nu} \eeq in terms of the Lyman limit frequency $\nu_{\rm LL}$
and $x_{\rm max}(n)$ corresponds to the comoving distance between $z$
and $z_{\rm max}(n)$ given by: \be 1+z_{\rm max}(n) = (1+z)
{\left[1-(n+1)^{-2}\right] \over(1-n^{-2})} \ .  \ee If the photon
emissivity, $\epsilon$ is homogeneous, equation \ref{Jax} can be
written as \citep{barkana05}, \be J_{\alpha}(z)=\frac{(1+z)^2} {4 \pi}
\sum_{n=2}^{n_{\rm max}} f_{\rm rec}(n)\int_z^{z_{\rm max}(n)} \frac{c
  dz'}{H(z')} \epsilon_\alpha(\nu_n', z')\ .
\label{Ja} 
\ee

\subsection{Gas temperature}
\label{sec:xray}

Once star formation has got underway a population of stellar remnants
will be produced capable of generating highly energetic X-rays.
Several candidate X-ray sources exist including X-ray binaries in
starburst galaxies, inverse Compton scattering of CMB photons from
relativistic electrons in supernova remnants (SNR) \citep{oh03a}, and mini-quasars.
X-rays may contribute to reionization, although constraints from the
unresolved soft X-ray background suggest that this is not the dominant
source of ionizing photons \citep{dijkstra04}.  More importantly, as
X-rays ionize hydrogen they deposit much of their energy as heat.
This X-ray preheating can easily be sufficient to heat the IGM above
the temperature of the CMB.  Although hard X-rays have a mean free
path comparable to the Hubble size, most of the heating turns out to
be done by soft X-rays ($E\lesssim 2$ keV) which can produce
significantly inhomogeneous heating at high redshifts
\citep{pritchard06}.
  
We calculate the X-ray heating following the model of
\citet{furlanetto06b}.  We model X-ray sources with a spectral
distribution function (number of photons per unit comoving volume per
unit time and frequency) that is a power-law with index $\alpha_S$:
\begin{equation}
\hat{\epsilon}_X(\nu)=\frac{L_0}{h\nu_0} \left(\frac{\nu}{\nu_0}\right)^{-\alpha_S-1},
\end{equation}
and the pivot energy is $h\nu_0=1\,\rm{keV}$.  We assume emission
within the band 0.1 keV to 30 keV and set the normalization constant
$L_0$ by requiring that the integrated power density $P=\int d\nu L_0
(\nu/\nu_0)^{-\alpha_S}$ is $3.4\times10^{40}
f_X\,\rm{erg\,s^{-1}\,Mpc^{-3}}$ when integrated between 0.1 keV and
30 keV, but with a highly uncertain constant factor $f_X$.  This
normalization is chosen so that, with $f_X=1$, the total X-ray
luminosity per unit SFRD (star formation rate density) is consistent
with that observed in starburst galaxies in the present epoch
\citep[see][for further details]{furlanetto06b}.  Typical values are
$\alpha_S=1.5$ for starbursts, $\alpha_S=1.0$ for supernovae remnants,
and $\alpha_S=0.5$ for the X-ray background generated by miniquasars.
These spectral indices are consistent with measured spectra of known
X-ray sources, though it is largely uncertain whether low-redshift
sources can be fully considered as a representation of source spectra
at high redshifts.

We link the total X-ray emissivity (number of photons per SFRD per
unit comoving volume per unit frequency and time) to the star
formation rate by
\begin{equation}\label{ehatX}
  \hat{\epsilon}_X(\bx,z,\nu)=\hat{\epsilon}_X(\nu)\left(\frac{\rm{SFRD(\bx,z)}}{\rm{M}_\odot \,\rm{yr}^{-1}\,Mpc^{-3}}\right).
\end{equation}
The X-ray number flux per unit frequency is then \be J_{X}(\bx,z,\nu)=
\int d^3x' {(1+z)^2\over 4\pi |\bx'|^2}
\hat{\epsilon}_X(\bx+\bx',\nu_n', z')e^{-\tau(z,\nu,\bx,\bx')}\ ,
\label{JX} 
\ee where again $|\bx'|$ is the comoving distance between $z$ and
$z'$, and $\nu'$ is the emission frequency at $z'$ corresponding to an
X-ray frequency $\nu$ at z \be \nu'=\nu {(1+z')\over (1+z)}\ .  \ee
The optical depth is given by 
\bea
\tau(z,\nu,\bx,\bx')\ &=&\ \int dl\ \left[n_{\rm
    HI}\sigma_{\rm HI}(\nu'')+\right.\\ \nonumber
&&\left. +n_{\rm HeI}\sigma_{\rm HeI}(\nu'')+n_{\rm HeII}\sigma_{\rm HeII}(\nu'')\right]\ , 
\eea
where the
integral is along the photon path in proper units between emission
($\bx+\bx'$) at redshift $z'$ and reception ($\bx$) at redshift $z$, 
while $\nu''$ is the frequency corresponding to $\nu$ at the redshift 
along the photon path. The cross-sections for ionization are 
calculated using the fits of \citet{verner96}.

Finally, the total rate of energy deposition per unit volume is \be
\epsilon_X(\bx,z)=\sum_i n_i \int_{\nu_{\rm th}}^\infty d\nu
\sigma_i(\nu) J_X(\bx,z,\nu) (h\nu-h\nu^i_{\rm th})\ ,
\label{eX}
\ee where $i=$HI, HeI, HeII, $n_i$ is the number density and
$h\nu_{\rm th}=E_{\rm th}$ is the threshold energy for ionization.  To
get the total heating rate, we multiply this by the fraction of energy
converted into heat $f_{\rm heat}$, obtained using the fitting
formulae of \citet{shull85}. We then evolve the gas temperature using
\begin{equation}\label{thistory}
  \frac{\ud T_K}{\ud t}=\frac{2T_K}{3n}\frac{\ud n}{\ud t}+\frac{2\epsilon_X f_{\rm heat}}{3 k_B n},
\end{equation}
where $n$ is the proper number density of all particles. In order
to evolve this equation through our simulation box, we set the 
initial condition for the gas temperature at $z=24.9$ to be
$T_K=14.43$ K.  The latter temperature is derived by assuming the
adiabatic cooling of the gas since recombination for our fiducial
cosmological model.  In setting this initial condition we also assume
that the gas is homogeneously cooled to this temperature at z=24.9.
Note that this is the same initial temperature as in our constant-temperature
case where we ignore fluctuations in X-ray heating of gas, among
others.

As $f_{\rm heat}$ depends on the free electron fraction in the IGM,
$x_e$, we must also evolve $x_e$ using 
\begin{equation}\label{xehistory}
\frac{\ud x_e}{\ud t}=\epsilon_X {f_{\rm ion}\over n E_{\rm th}}\ ,
\end{equation}
where $f_{\rm ion}$ is the fraction of energy converted into
ionizations which also depends on $x_e$ (note that $\epsilon_X$ also
depends on $x_e$ through $n_i$ in equation~\ref{eX}). 
Note that this term is quite important since otherwise
we get $x_e\sim 0$ and $f_{\rm heat}\sim 0$ in the IGM from UV ionization while 
$f_{\rm heat}\sim 0.01-0.1$ when $x_e\sim 10^{-8}-10^{-4}$ instead of 0. 
We neglect primary ionizations from X-rays since secondary ionizations dominate at large
radii from halos and UV ionizations dominate at small radii.
In making the above
calculation, since $f_{\rm heat} \propto x_e^{1/4}$ \citep{shull85},
we approximate it by only considering the ionization of
hydrogen and by setting $E_{\rm th}=13.6$ eV.  Both recombinations and
corrections from helium are typically not important for calculating
$x_e$, which stays small over the redshift range of interest.  Figure
\ref{aver_xi} shows the evolution of $x_e$ in the IGM for the
simulation, which remains small at all times.  Note that $x_e$ is
defined in the neutral IGM outside of fully ionized HII regions.


\section{The 21-cm signal: simulations}

In order to obtain the 21-cm brightness temperature from simulations,
we basically need to apply equation \ref{t21}. Both $f_{\rm HI}$ and
$\rho_b$ are already obtained by the simulation as these properties are calculated with the
evolution of  the dark matter distribution. On the other hand,
the spin temperature, $T_S$, is calculated at the
post-processing stage since the radiative transfer calculations of the initial run do not take
into account the \Lyman photons and X-ray heating.  Usually it is
assumed that $T_S>>\tcmb$ (e.g. the number density of hydrogen atoms
in the triplet level is saturated) so that one does not need to worry
about the spin temperature contribution to the 21-cm signal
\citep{mellema06}.  However, this assumption should only be safe for
$z<10$ so that one needs to consider the evolution of the spin
temperature for a proper treatment of the 21-cm signal at the higher
redshifts provided by this simulation. Moreover, fluctuations in the
\Lyman coupling and X-ray heating, may be important at high redshifts
\citep{barkana05,pritchard06,semelin07}.  Therefore, we present here a
full calculation of these fluctuations on the high redshift signal
probed by the simulation.

\subsection{Coupling parameters}

Calculation of the collisional coupling parameter from equation
\ref{yc} is straightforward. In this case we can easily include
perturbations due to fluctuations in $n_{\rm HI}$ and $n_e$.  In order
to obtain the radiative coupling parameter, $y_\alpha$ we need to
determine the comoving photon emissivity directly from the simulation
using: \be \epsilon_\alpha(\bx,\nu, z) =
SFRD(\bx,z)\epsilon_b(\nu)\ , \label{emit} \ee where $SFRD(\bx,z)$ is
again the star formation rate density from the simulation (in terms of
the number of baryons in stars per comoving volume and proper time)
and $\epsilon_b(\nu)$ is the spectral distribution function of the
sources (defined as the number of photons per unit frequency emitted
at $\nu$ per baryon in stars). Note that we are assuming that stars
dominate over mini-quasars for the radiative coupling. We consider
separately the spectral distribution function from Pop II stars
\citep{leitherer99} and Pop III stars (\citealt{bromm01}, but see also
\citealt{barkana04}).  We then apply directly equation \ref{Jax} to
the simulation boxes. We can speed up this calculation by noting that
the integral can be written in terms of a convolution between the SFRD
and a specified kernel.

\begin{figure*}[!t]
\centerline{\hspace{-0.0cm}\includegraphics[scale=0.45]{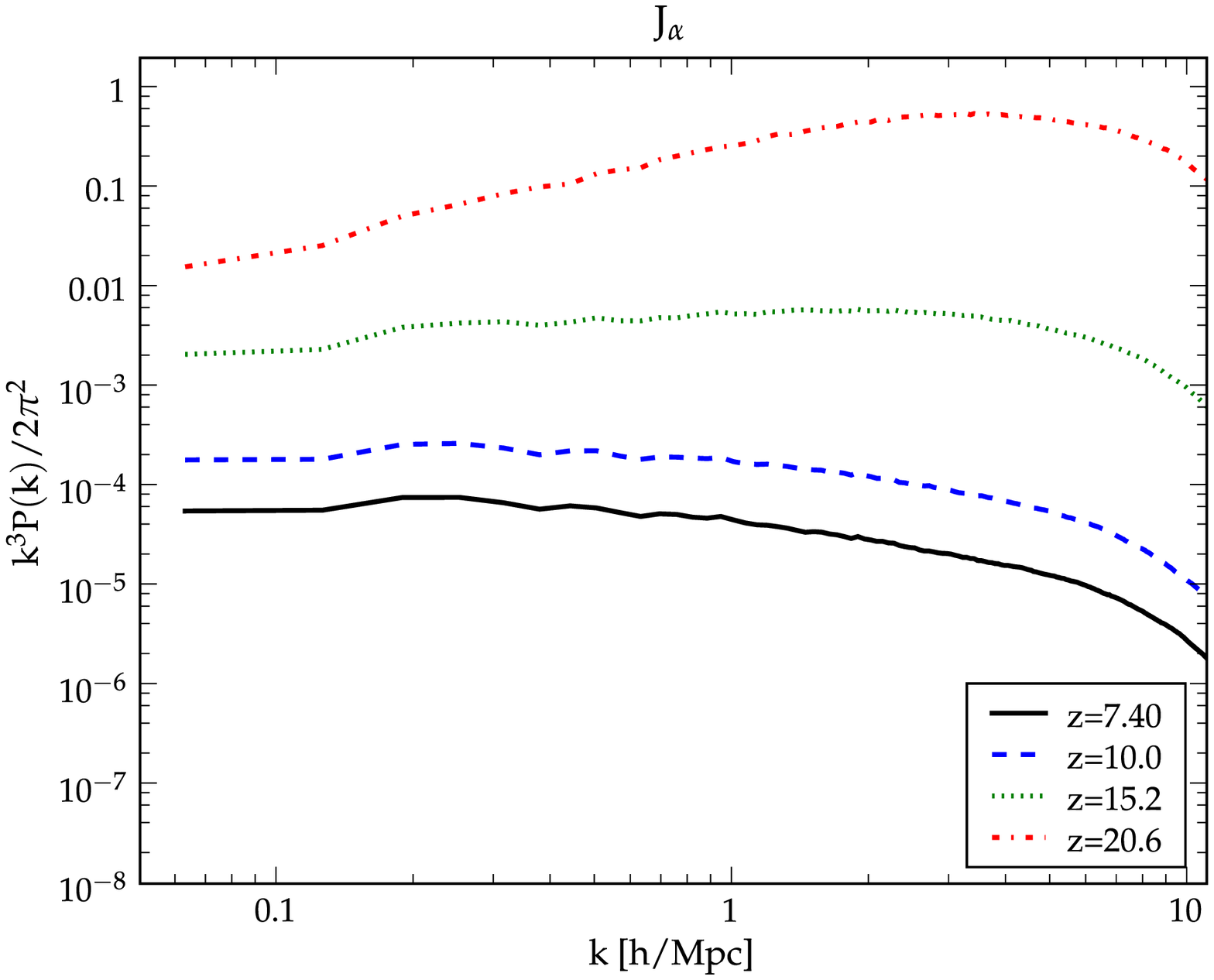}\hspace{0.0cm}  \includegraphics[scale=0.45]{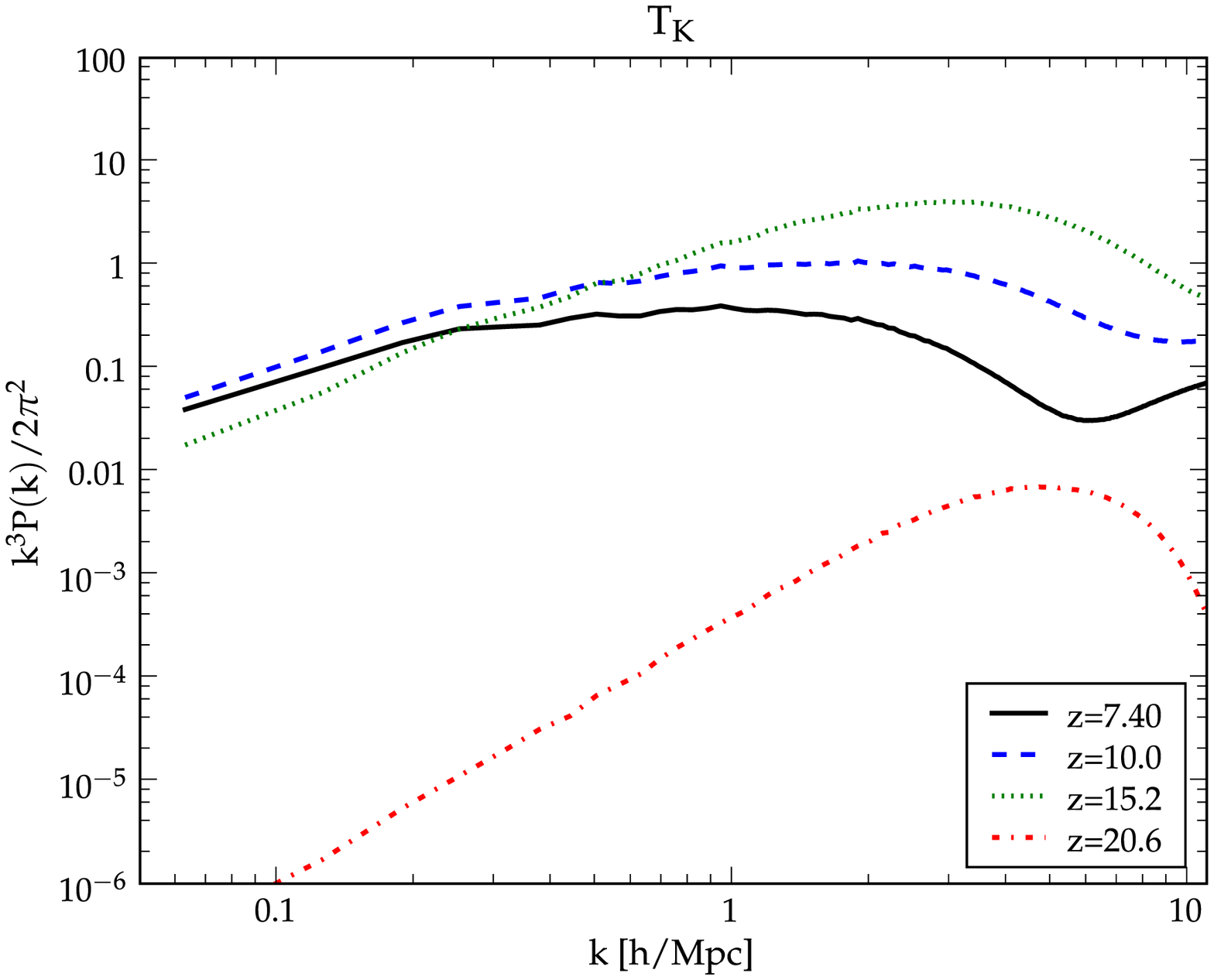}}
\caption{Left: Power spectra of the \Lyman flux at several redshifts
divided by the square of the mean flux. 
Right: power spectra of the gas temperature due to x-rays divided by the
average squared at each redshift.}
\vspace{0.5cm}
\label{JaTK}
\end{figure*}

Figure \ref{JaTK}, left panel, shows the power spectrum of the $J_\alpha$
fluctuations from the simulation. The power spectrum is dominated by
large scale fluctuations at low redshifts, since most of the photons
propagated through the entire box and fluctuations represents a small
percentage of the overall \Lyman photon flux. At large redshift
(around $z\simeq20$), most of the \Lyman photons have not had time to
propagate very far from halos and cluster around halos in 0.2-0.3
$h^{-1}$ Mpc bubbles. However, these bubbles are highly clustered and
have large bias factors related to the dark matter density field. The
strong clustering of these small bubbles around first-light sources
still leave a \Lyman intensity power spectrum that is strongly
clustered even at high redshifts.  Our measurement directly by
post-processing the star-formation rate in our numerical simulation to
extract properties of gas physics and 21-cm brightness temperature
shows that there is no Poisson or shot-noise component in the \Lyman
intensity fluctuations. Analytical models that motivate detections of
first-light galaxies from the 21-cm background have suggested a large
shot-noise for the \Lyman intensity background fluctuations
\citep{barkana05}, but we do not see such a component in our
simulations to the extent that we can separate the \Lyman intensity
power spectrum at redshifts $z \sim 20$. While we cannot comment on
the appearance of a shot-noise even at higher redshifts or smaller scales, 
we hope to return to the general issue of detecting signatures of the highest
redshifts galaxies in the 21-cm background in an upcoming paper by
fully taking into account the instrumental systematics and foregrounds.

Finally the \Lyman coupling is determined through equation
\ref{yalpha} (there will be extra fluctuations due to the $S_\alpha$
dependence on $T_K$ and the matter density distribution).  Analysis of
the simulation shows that on average $y_\alpha>1$ for $z\lesssim 17$
signaling the approach of the spin temperature to the gas
temperature. Moreover, the \Lyman coupling dominates over collisions
up to $z\sim 22$ when $y_c\sim 10^{-2}$.

\subsection{Gas temperature}

As described in Section \ref{simulation}, the gas temperature provided
by the simulation is essentially due to the virial temperature, given
by the velocity dispersion of each particle. This effectively sets the
gas temperature to $T\gtrsim 10^4$ K within the virial radius of
halos.  However, most of the heating is restricted to the high
density, ionized regions, while the neutral IGM continues to cool
adiabatically. This means that most of the 21-cm signal would be seen
in absorption even at the low redshifts when reionization is
well underway.

X-ray heating on the other hand should heat the neutral IGM
above the CMB temperature fairly easily and needs to be taken into
account for a proper treatment of the 21-cm signal. We need therefore
to include X-ray heating as part of our post-processing to predict the
spin temperature of gas and to calculate the brightness temperature of the 21-cm
signal.  Due to the clear separation between the ionized and neutral regions, we can consider
the evolution of both heating mechanisms separately.

\begin{figure}[!t]
\hspace{-0.4cm}
\includegraphics[scale=0.43]{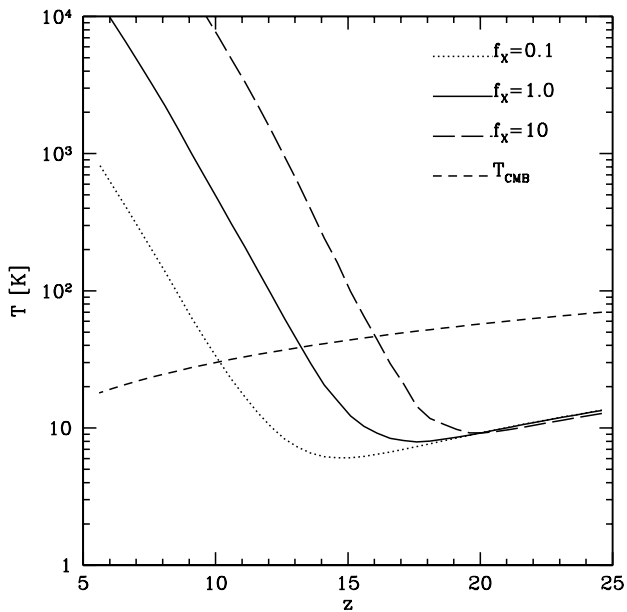}
\caption{Evolution of gas temperature due to X-rays for several $f_X$
  (normalization of the X-ray luminosity).}
\label{temp_fx}
\end{figure}

In order to calculate heating due to X-rays, we basically follow the
procedure outlined in section \ref{sec:xray}, using the star formation
rate density provided by the simulation. In order to use the same
Fourier transform technique to speed up the temperature fluctuation
calculation, we assumed the density of our species to be spatially constant and
equal to the average in the computation of the optical depth. This
approximation will reduce the optical depth around the halo, but as
one get further away our optical depth should converge to the real
one. We therefore probably underestimate the temperature in the halos
and overestimate in the neighboring regions outside. For the spectral
distribution function we assume $\alpha_S=1.5$ (starbursts).  
Figure \ref{JaTK}, right panel, shows the dimensionless power spectrum of 
the gas temperature due to X-rays from the simulation. Note that these
fluctuations are only relevant to the 21-cm signal as long as the gas
temperature is not much higher than the  CMB temperature.
Thermal histories are also plotted in Figure \ref{temp_fx} for $f_X=0.1$, $1.0$,
and $10.0$.  These indicate that X-ray preheating can indeed heat the
gas above the CMB temperature at the redshift range important for 21-cm observations,
justifying the assumption that $T_S\gg T_{\rm CMB}$ at redshifts $z <
10$.  Clearly, though, there is considerable uncertainty in what the
exact thermal history is expected to be.  Hereafter, when we calculate the 21-cm
brightness temperature and its anisotropy, we will assume
$f_X=1.0$. The analytical result we present here is also matched to
the same X-ray intensity.

To highlight the reason why we consider X-ray heating as an
inhomogeneous process, in Fig.~7, we plot the volume filling factor of
our simulation calculated by taking the ratio $\sum_i
(4/3\pi\lambda^3)/V_{\rm simul}$ where $\lambda$ is the mean free path
around each X-ray source identified in the simulation with volume
$V_{\rm simul}$.  While photons with energies around 100 eV that are
primarily responsible for heating propagate rapidly and fill the box
by $z \sim 10$, the volume filling factor is below 1 around $z \sim
15$. This demonstrates that fluctuations in the heating of the gas by
X-rays will be important around these redshifts.

\begin{figure}[!t]
\hspace{-0.6cm}
\includegraphics[scale=0.52]{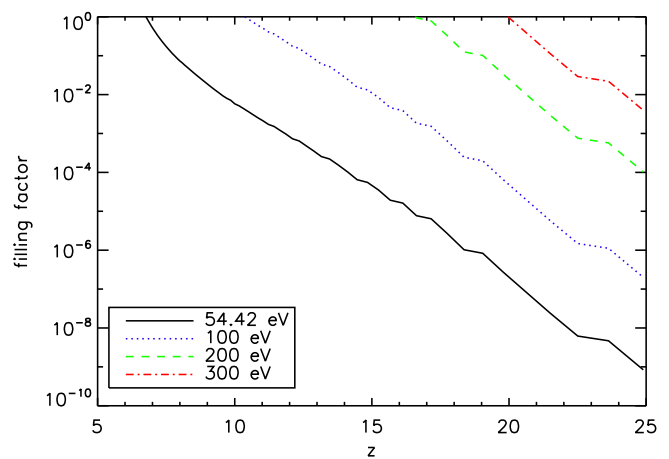}
\caption{ Volume filling factor of the X-ray radiation in the
  simulation as a function of redshift for several different values of
  the photon energy.  While a photon with an energy such as 100 eV,
  that dominates X-ray heating, fills up the whole volume by $z \sim
  10$ (ie. mean free path is larger than the simulation box length), 
  at $z \sim 15$, the volume filling factor
  remains below 1 and X-ray heating at the onset of reionization is
  inhomogeneous.}
\label{vol_x}
\end{figure}

\begin{figure}[!t]
\hspace{-0.55cm}
\includegraphics[scale=0.48]{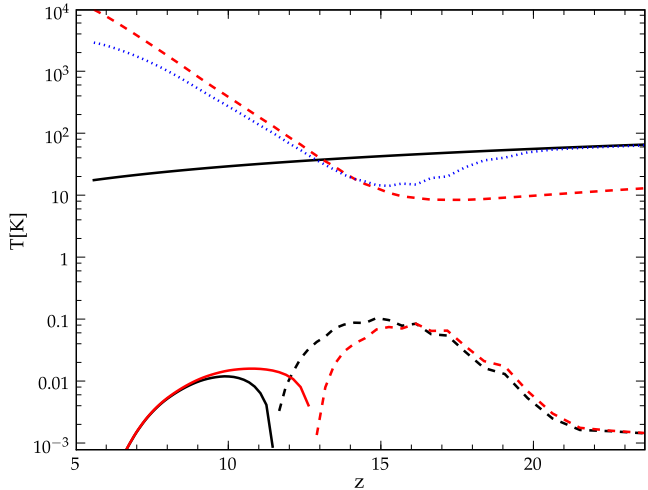}
\caption{Temperature of the CMB (black solid line), gas (red dashed
  line) and spin (blue dotted line) as a function of $z$, in our
  simulation, in which we have included x-ray heating, collisional and
  radiative coupling. The lower, solid black line shows the average
  brightness temperature when all fluctuations are taken into account,
  while the dashed one shows the absolute value (since it is negative
  at these redshifts). Lower, red (grey) line shows the brightness
  temperature when only fluctuations in the matter density and
  ionization fraction are included.}
\label{temp_sim}
\end{figure}

\subsection{Brightness temperature}

Finally, implementing equation \ref{t21}, we can calculate the 21-cm
brightness temperature for the simulation.  Figure \ref{temp_sim}
shows the average evolution of this brightness temperature, together
with the CMB temperature, gas temperature, and the spin temperature.
Note that, as already pointed out, the spin temperature decouples from
the CMB at $z\lesssim 17$ when $y_\alpha>1$. The average 21-cm signal
is clearly non-negligible at high redshifts. Moreover the transition
from emission to absorption depends on whether X-ray fluctuations are
included. This is because regions with very low ionization fraction
are still cold in spin temperature at low redshifts relative to the CMB and are therefore seen in
absorption. 

\begin{figure}[!t]
\hspace{-0.55cm}
\includegraphics[scale=0.48]{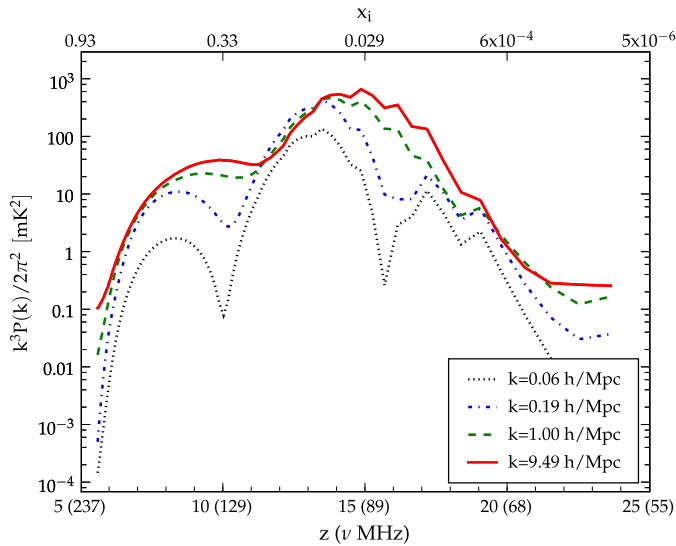}
\caption{Evolution of the brightness temperature power spectrum with redshift.}
\label{t21_z}
\end{figure}

Figure \ref{t21_z} shows the evolution of the 21-cm signal with
redshift for a set of $k$ values.  Again we can see that the signal
strength actually increases for $z>12$.  The evolution of the 21-cm
brightness fluctuations resemble strongly the average 21-cm brightness
temperature. The
fluctuations increase up to $z=15$ as the coupling of the spin
temperature to the gas temperature increases, then diminish during
the absorption-emission transition down to $z=11$, increasing or reaching a plateau in
the $T_S >> T_{CMB}$, regime and then falling again at $z=8-9$ when the
ionization fraction reaches very high values and the universe fully reionizes.

The evolution of the power spectrum on large scales is qualitatively similar to 
that calculated in \citet{pritchard08} showing three peaks resulting from the
periods where ionization, temperature, and Ly-alpha fluctuations respectively
come to dominate. We note that the dip at $x_i\sim0.3$ occurs because on large scales
the gas temperature is quite close to the CMB temperature at this redshift.

\begin{figure*}[!t]
\centerline{\hspace{1cm}\includegraphics[scale=0.6]{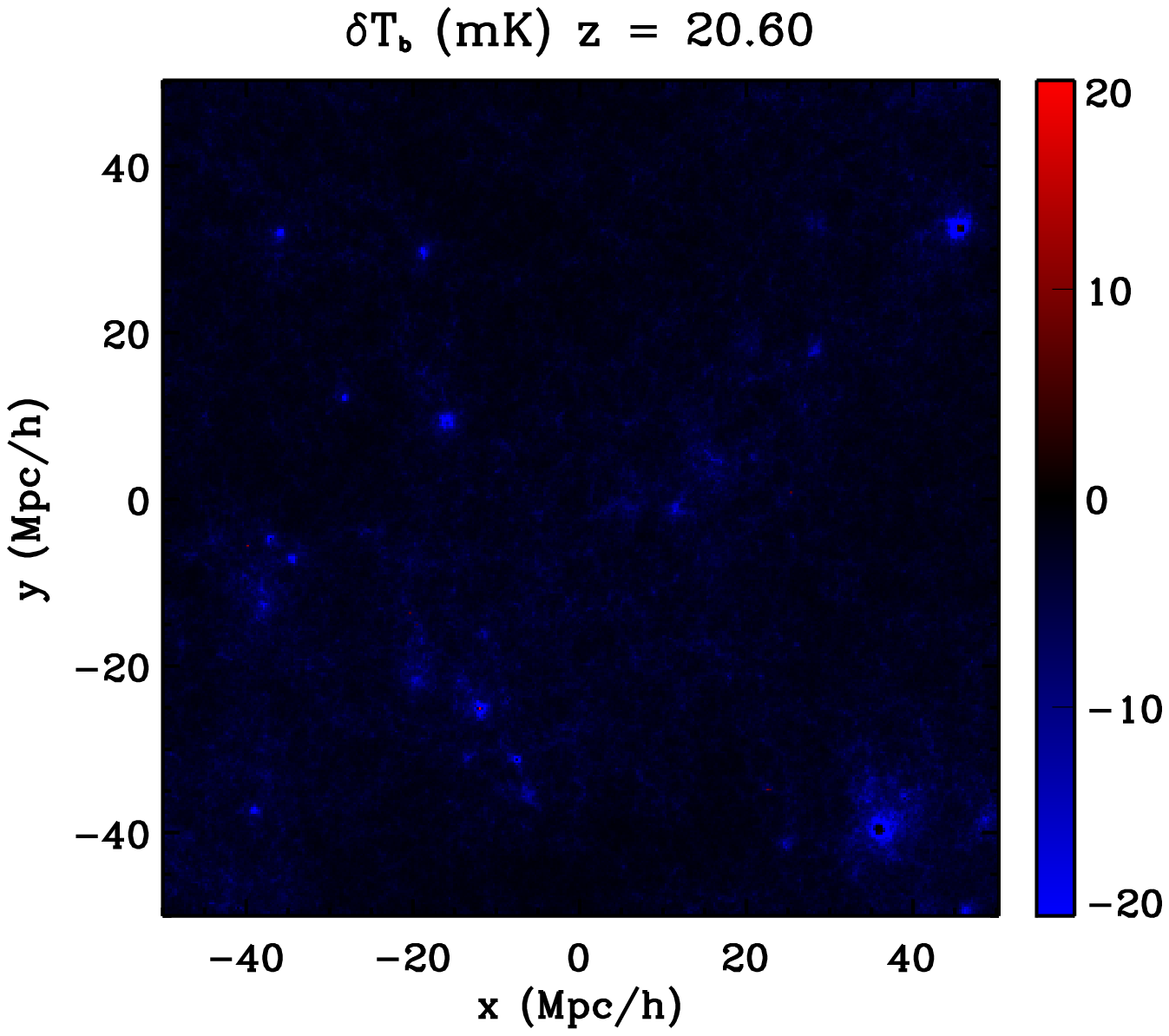}\hspace{-2cm}  \includegraphics[scale=0.6]{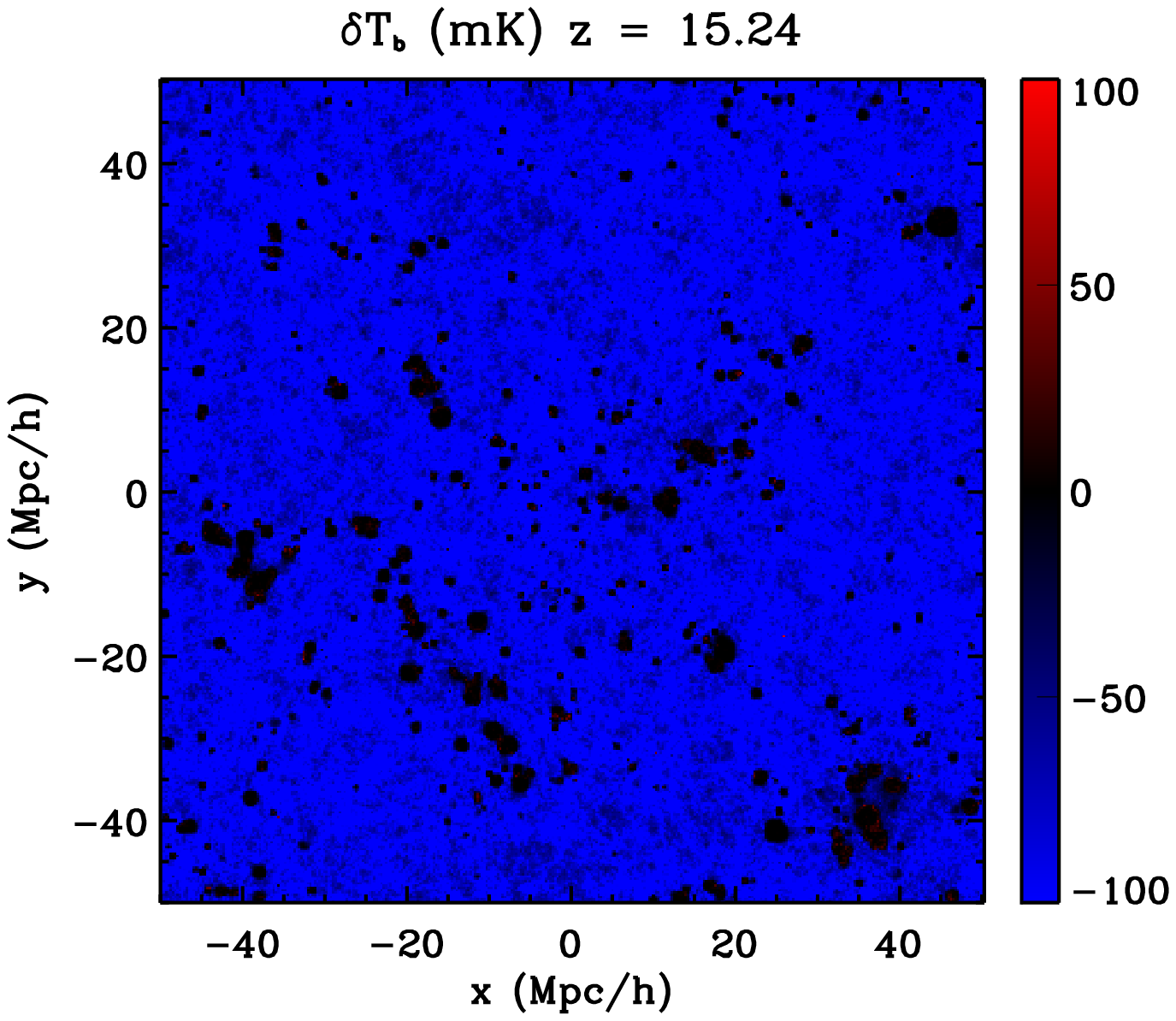}}
\centerline{\hspace{1cm}\includegraphics[scale=0.6]{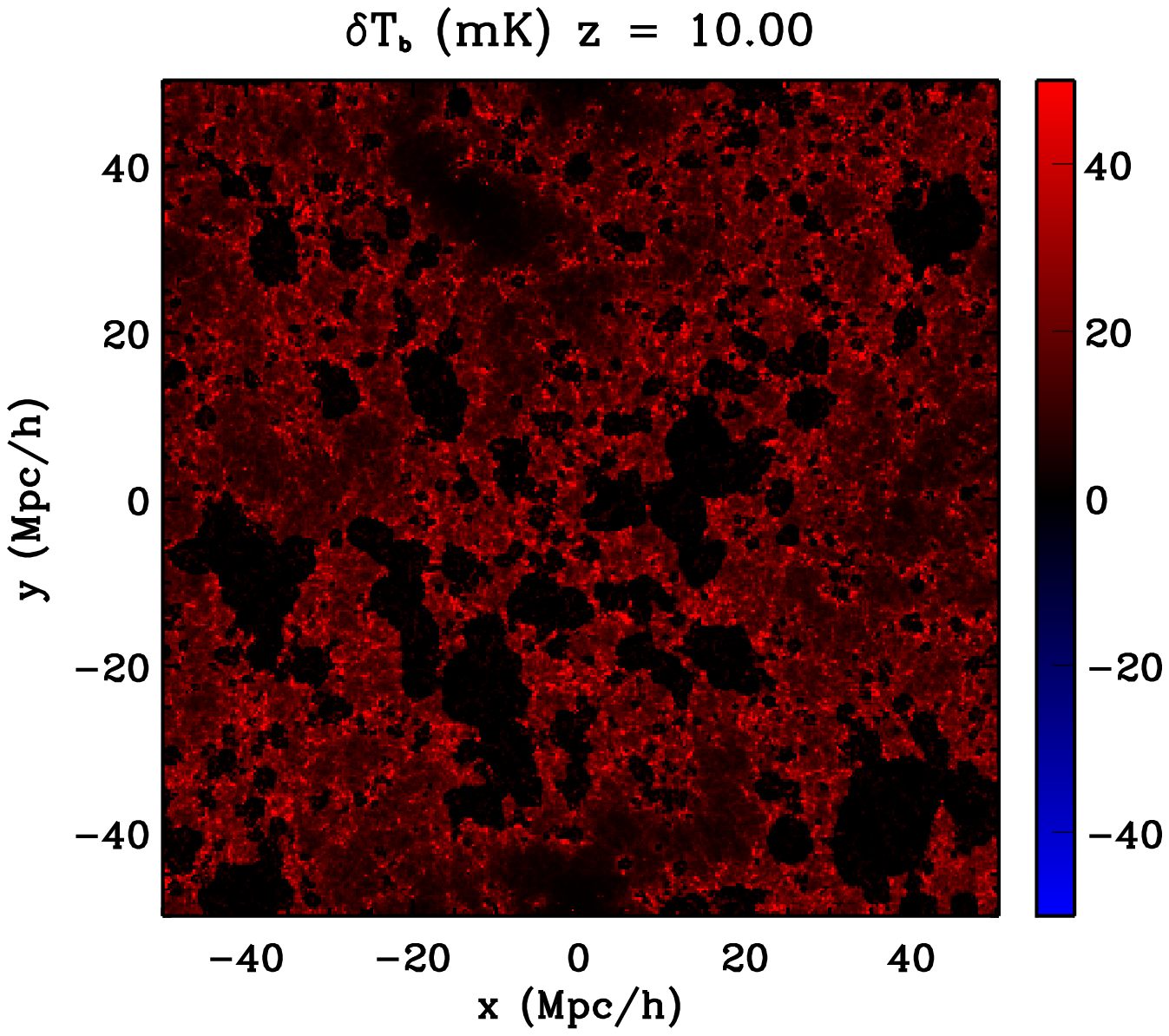}\hspace{-2cm}  \includegraphics[scale=0.6]{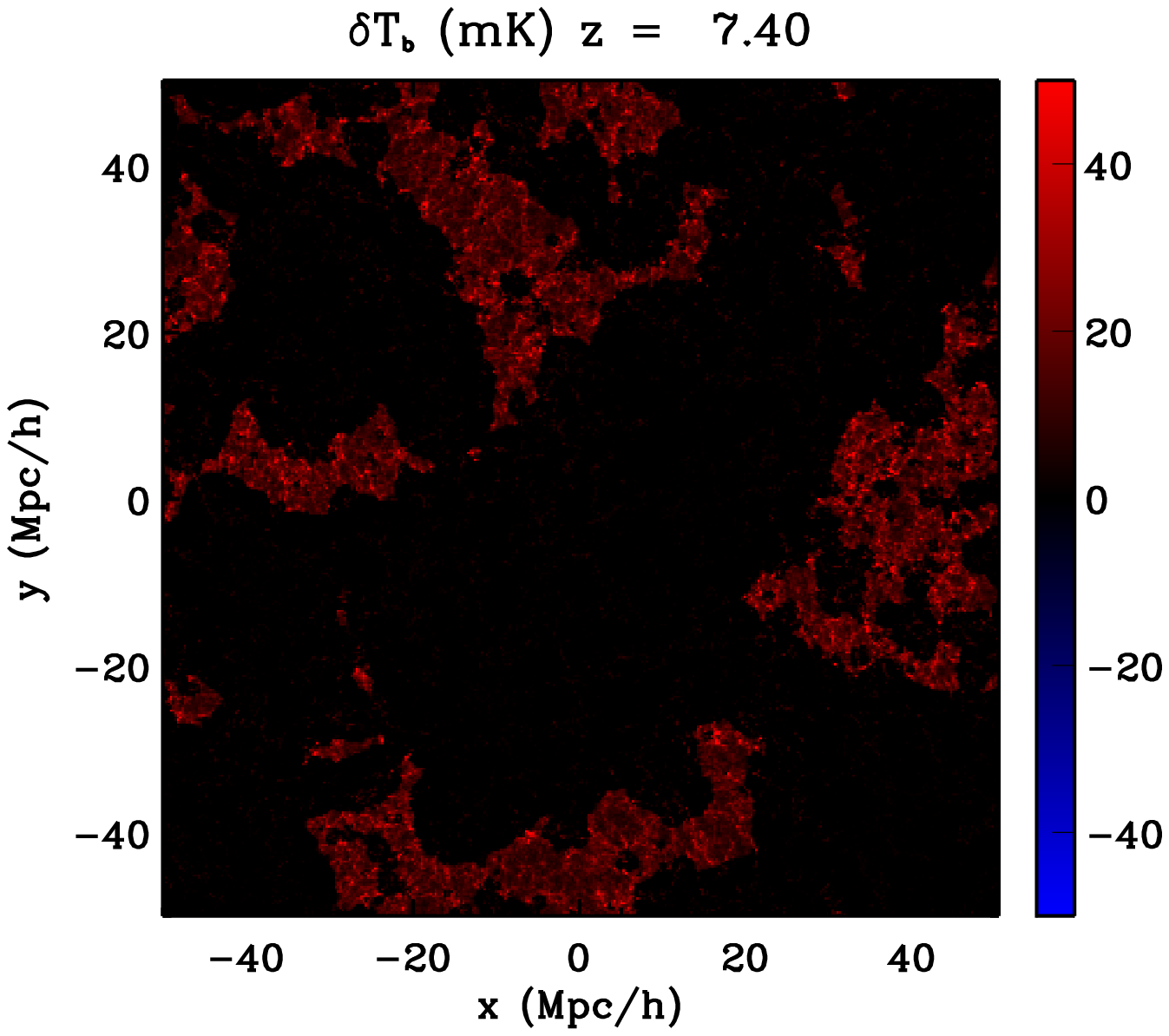}}
\caption{Maps of the 21-cm brightness temperature from the simulation,
  including all fluctuations, at redshifts $z=20.6,15.2,10.0,7.4$, as
  labeled, corresponding to $\bar{x}_i=0.0002, 0.03, 0.35, 0.84$.}
\label{map_T21}
\end{figure*}

Figure \ref{map_T21} shows the brightness temperature for the same
maps as in figure \ref{map_xi}, including fluctuations in the matter
density, ionization fraction, \Lyman coupling and X-ray heating.
Note that some of the features from the reionization maps are somewhat
smeared in the brightness temperature maps due to the convolution with
the density.  At $z\sim 15$ perturbations in \Lyman and X-ray heating
are already non-negligible and need to be properly taken into
account when making 21-cm predictions. 

\begin{figure*}[!t]
\centerline{\hspace{-0.1cm}\includegraphics[scale=0.9]{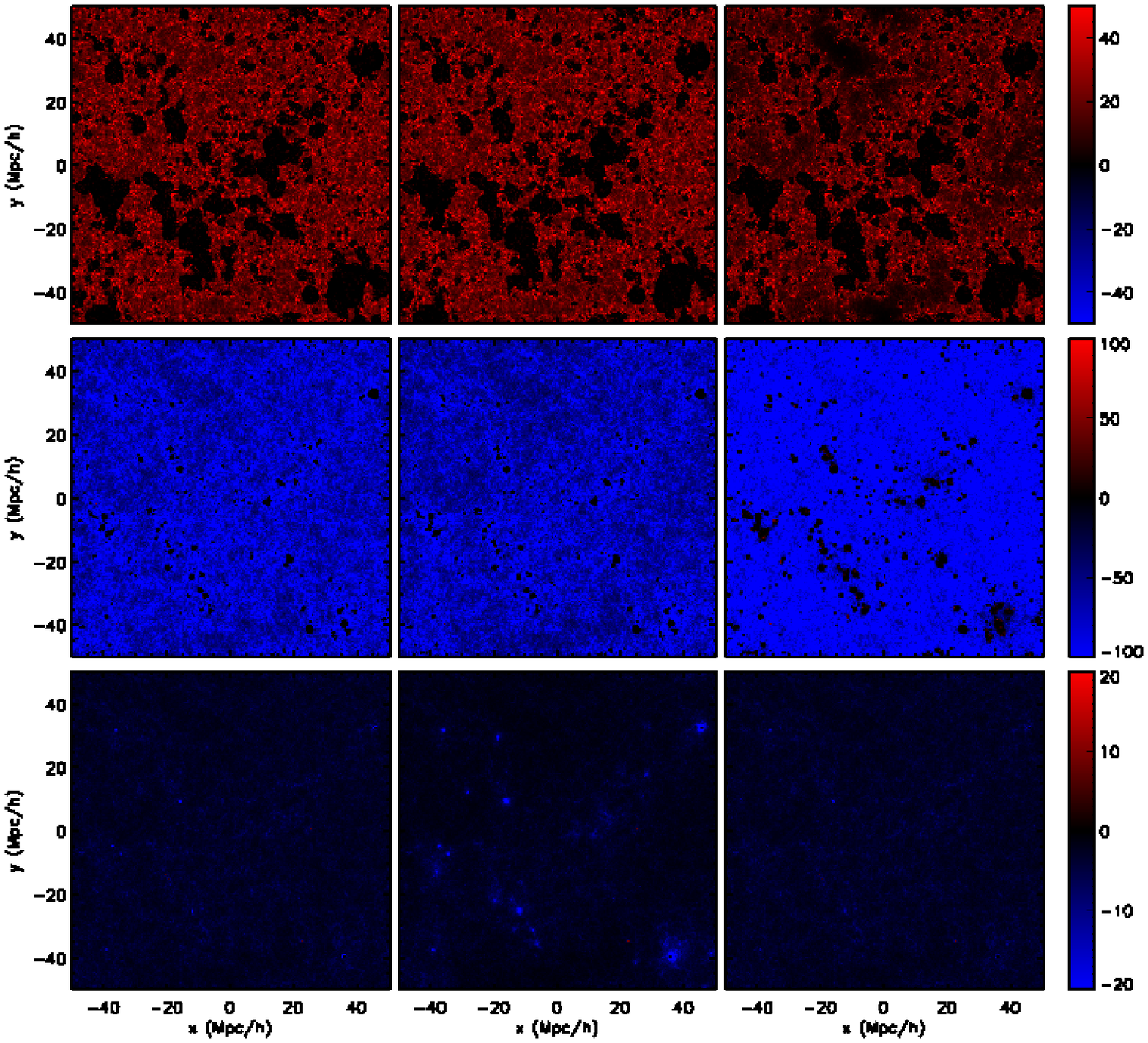}}
\caption{Maps of the 21-cm brightness temperature. Top:
  $z=10$. Middle: $z=15$. Bottom: $z=20$. Left: fluctuations in
  density and ionization fraction with homogeneous X-ray heating and
  \Lyman radiation field.  Center: Fluctuations in the \Lyman
  background added, but homogeneous X-ray heating. Right: fluctuations
  in X-ray heating, but uniform \Lyman radiation field.}
\vspace{0.4cm}
\label{map_T21_pert}
\end{figure*}

Figure \ref{map_T21_pert} shows the 21-cm signal at $z=20, 15$ and
$z=10$ from top to bottom respectively, where we separately include
fluctuations in following quantities: matter density plus ionization fraction (left
panel), matter density plus ionization fraction plus \Lyman coupling (middle panel), and
matter density plus ionization fraction plus X-ray heating (right panel).  While at $z \sim 20$
fluctuations in the \Lyman intensity field are important, at $z \sim 15$,
features in the brightness temperature are sensitive to whether we
model X-ray heating as an inhomogeneous source or whether we consider
X-ray heating, wrongly, as uniform.  This is the redshift in our
simulation at which , due to X-ray heating, regions are beginning to
be first detected in emission of the 21-cm brightness temperature
instead of absorption as at higher redshifts.

In terms of the inhomogeneous sources of the 21-cm brightness
temperature, fluctuations in \Lyman background makes the biggest
difference at $z \sim 20$ as can be seen from the bottom middle
panel. Here, the locations with largest absorption signal in the 21-cm
background is associated with first-light sources that have just
formed first stars.  A comparison of the middle panel of
Fig.~\ref{map_T21_pert} at $z=20$ and the right panel at
$z=15$ shows that regions that are first dominating in emission due to
X-ray heating at $z\sim 15$ are mostly the same regions that were
first brighter in absorption at $z\sim 20$ due to \Lyman coupling.  At
lower redshifts, however, during partial reionization, fluctuations in
the brightness temperature are dominated by fluctuations in the
ionization fraction modulated by the density field inhomogeneities.
Thus, at lower redshifts $z < 10$ that will be targeted by the
first-generation 21-cm interferometers, one can mostly ignore the
effects associated with inhomogeneous X-ray heating or anisotropies in
the \Lyman coupling.  This becomes clear when we compare the
predictions related to the 21-cm brightness temperature anisotropy
power spectrum with those related to analytical models in the next
Section.

\section{Overall Comparison of the 21-cm signal: Simulation and Analytical Model}

Here we compare the power spectrum of the 21-cm signal from the
simulation to a fast analytical power spectrum generator.  The
calculation of the analytical power spectra will basically follow the
procedure outlined in \citet{pritchard06}, using the model for the
ionization fraction discussed in section \ref{sec:models}, and we
refer the reader to \citet{pritchard06} for further details.

When considering fluctuations in the 21-cm brightness temperature
arising from variations in ionization and density, the correlation
function of the brightness temperature can be expressed as \bea
\label{21corr}
\xi_{T_b T_b}&\equiv& \left<(\delta T_b - \bar{\delta T_b})(\delta T_b
- \bar{\delta T_b})\right>\\ \nonumber &=&T_c^2\left[\bar{f}_{\rm
    HI}^2\xi_{\delta\delta}+\xi_{x_ix_i}(1+\xi_{\delta\delta})-\xi_{x_i\delta}(2\bar{f}_{\rm
    HI}-\xi_{x_i\delta})\right] \eea and we use the procedure
described in section \ref{sec:models} to calculate
$\xi_{\delta\delta}$, $\xi_{x_ix_i}$ and $\xi_{x_i\delta}$. Note that
we assume gaussianity in this calculation but take into account the 
Gaussian terms from the 4-point function (see the appendix for a discussion
of the effect of the non-Gaussian terms in equation \ref{21corr}).
However, variations in the ionization fraction and density are not the
only sources of 21-cm brightness temperature fluctuations and we include
anisotropies in X-ray heating and \Lyman coupling separately (by following the
procedure in \citealt{pritchard06}) so
we can study how predictions change with whether one considers heating
to be uniform, for example.  We also follow the same procedure
with the analytical calculation by adding various sources in step for
easy comparison with numerical simulations.

In order to calculate the gas temperature (due to heating by X-rays),
we need to obtain the analytical star formation rate and use it in
equation \ref{ehatX}. We model the star formation rate as tracking the
collapse of matter, so that we may write the star formation rate per
(comoving) unit volume
\begin{equation}\label{theorySFR}
{\rm SFRD}=\bar{\rho}^0_b(z) f_{*}\frac{\ud }{\ud t}f_{\rm{coll}}(z).
\end{equation}
where $\bar{\rho}^0_b$ is the cosmic mean baryon density today.  This
formalism is appropriate for $z\gtrsim10$, as at later times star
formation as a result of mergers becomes important.  For the
analytical calculation, we do not distinguish between Pop II and Pop
III stars and so use a value of $f_{*}=0.1$, appropriate for Pop II
stars, which dominate star formation at lower redshifts.  While these
parameters have not been fitted to the simulation data, the
star-formation rates from theory and simulation agree quite well.
Fourier transforming the correlation function, $\xi_{T_b T_b}$, in
equation \ref{21corr} yields the desired power spectra. By first
generating the correlation functions and then Fourier transforming we
avoid having to consider the power spectrum convolution for the
$\xi_{x_ix_i}\xi_{\delta\delta}$ and $\xi_{x_i\delta}\xi_{x_i\delta}$
terms.

\begin{figure*}[!t]
\centerline{\includegraphics[scale=0.5]{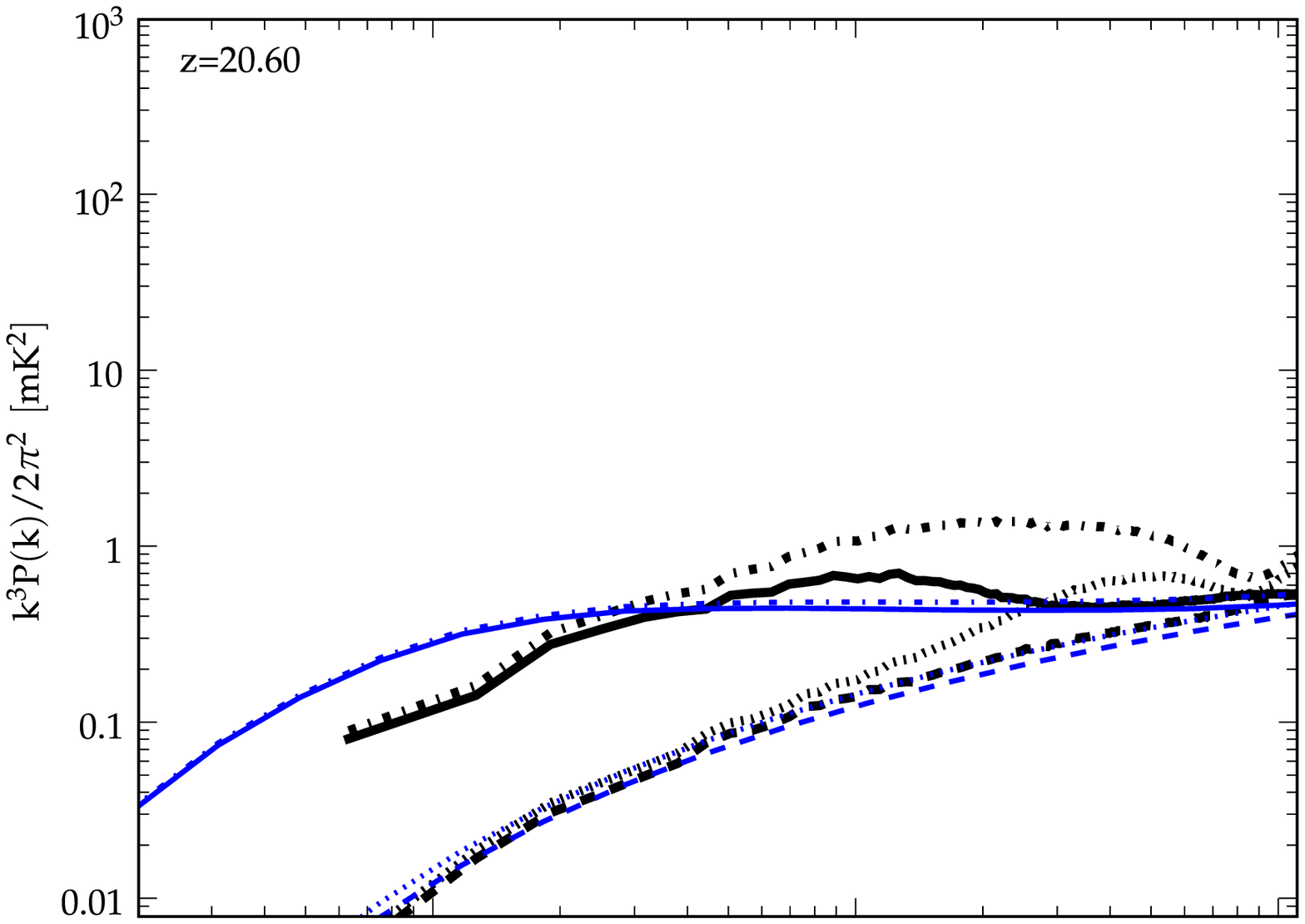}\hspace{-2.3cm} \includegraphics[scale=0.5]{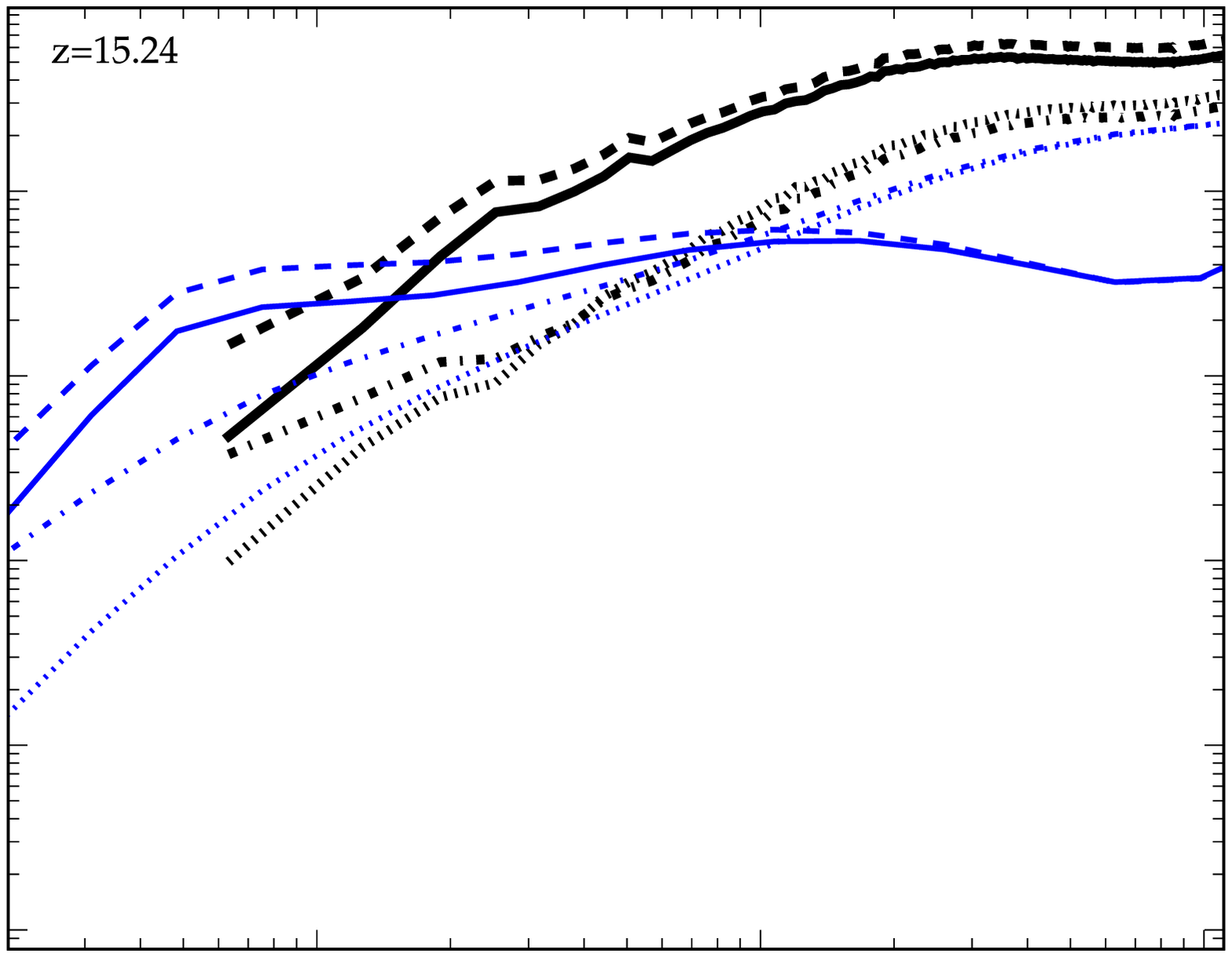}}
 \vspace{-1.4cm}
 \centerline{\includegraphics[scale=0.5]{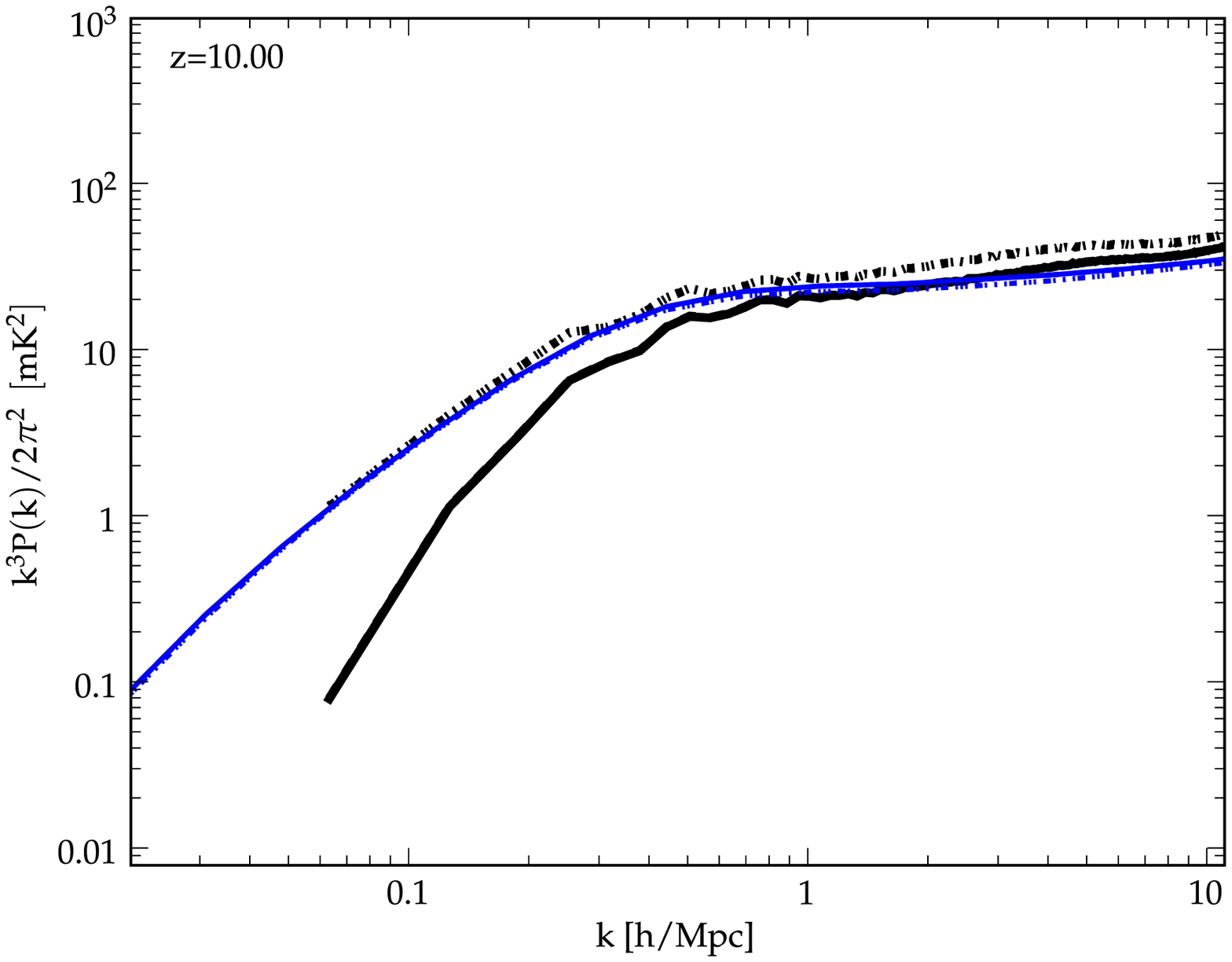}\hspace{-2.3cm} \includegraphics[scale=0.5]{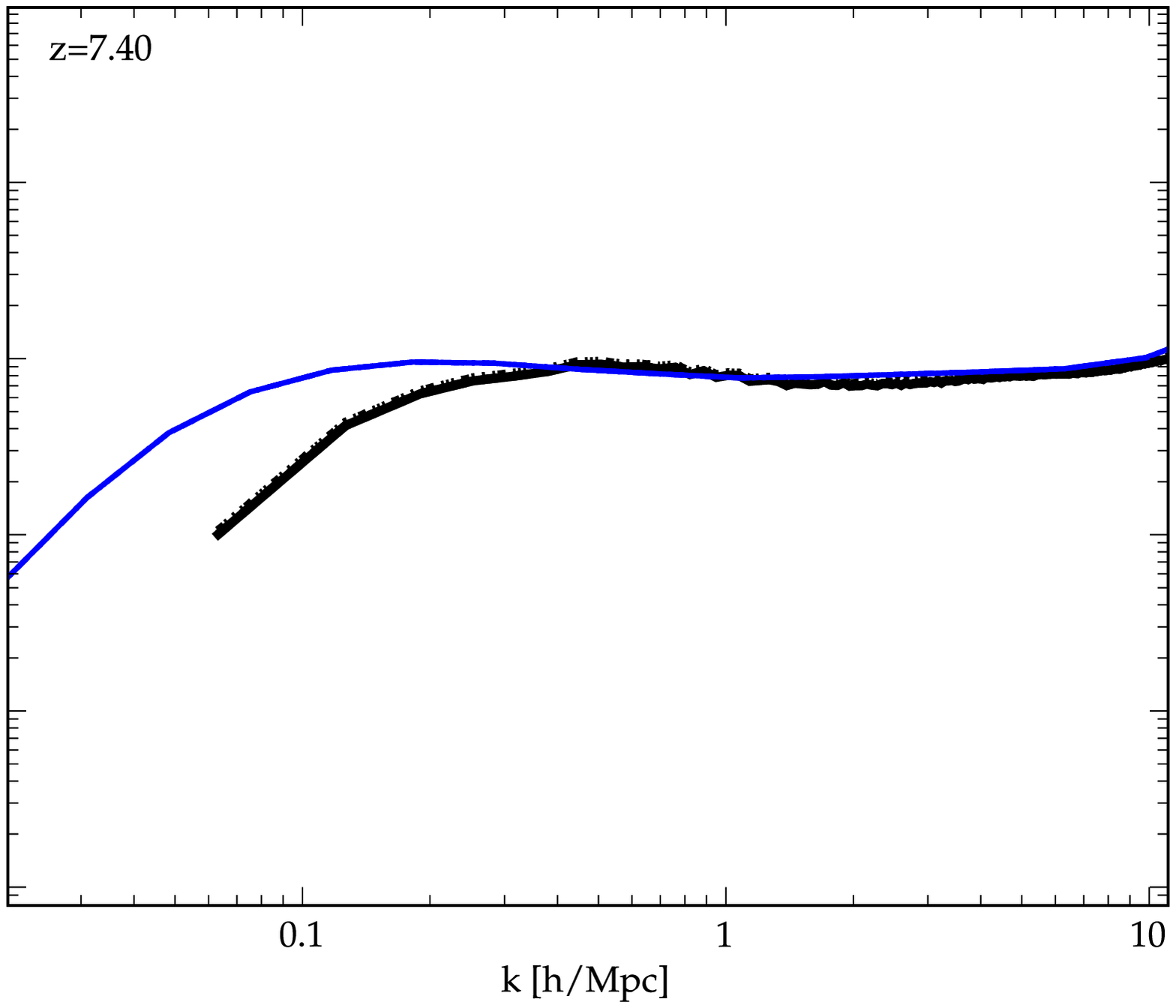}}
\caption{The power spectra of the 21-cm brightness temperature for
    the simulation (thick black curves) versus theory (thin blue
    curves).  Solid: all fluctuations included; dotted: fluctuations
    in the matter density ($\rho_m$) plus ionization fraction ($x_i$);
    dot-dashed: fluctuations in $\rho_m$ plus $x_i$ plus \Lyman;
    dashed: perturbations in $\rho_m$ plus $x_i$ plus X-rays.}
\vspace{0.4cm}
\label{T21power}
\end{figure*}

In figure \ref{T21power} we show a comparison of the brightness
temperature power spectrum between the analytical calculation and
numerical simulations.  Previously we have commented on the comparison
between the analytical model prediction for the ionization fraction
power spectrum (relative to the density field power spectrum) and the
numerical simulation for the same quantity (Fig.~3 and Section~2.2).
We also expect the 21-cm prediction which uses
only the ionization fluctuations and the density field as
inhomogeneous sources to agree with the analytical calculation to the
same extent that the two agreed previously. Thus, any differences
beyond that of Fig.~3 in Fig.~\ref{T21power} between simulations and
the analytical model are due to differences in the two prescriptions
related to fluctuations in the \Lyman intensity field and X-ray
heating.

At low redshifts, we note that the two predictions agree very well to
the extent that we can trust our simulations.  At $z\sim 7$, the
difference at $k \sim 0.1$ h Mpc$^{-1}$ is most likely to be that of
the finite volume of the box and the cut-off associated with
large-scale bubbles that have grown beyond the volume of the
simulations, since we find less power in the simulation compared to
that of the analytical model (see figure \ref{reion_Pk}). At $z \sim 10$ the agreement between the
simulation and analytical curves corresponding to density plus ionization
fraction fluctuations (dotted lines) is very good. The difference in the solid
lines is because x-ray perturbations are actually important at these redshifts.
When we include fluctuations in x-rays, the very
large, neutral scales ($k<0.4$) are still "cold" at $z=10$ (close to the CMB
temperature) so we get a reduction in the 21-cm perturbations (this can also
be seen in Figure \ref{t21_z}).
The differences between the two prescriptions become more obvious
at $ z\sim 15$ and
higher. At these high redshifts, as we have already noted, different
sources of fluctuations make different contributions to the 21-cm
brightness temperature fluctuations.

At $z \sim 15$, analytical calculations suggest a characteristic scale
for 21-cm brightness temperature fluctuations with a shot-noise like
power-spectrum at $k > $ a few h Mpc$^{-1}$.  This 21-cm power
spectrum is dominated by fluctuations in the X-ray heating background
as can be established from a comparison between the dotted and dashed
lines related to the analytical calculation at $z\sim15$.  At this
redshift, the analytic model shows a 
considerable deficit of power on small scales when compared to the simulation and
inhomogeneous X-ray heating is included.  This appears to be a result of the 
analytic model assuming a tight cross-correlation between temperature and density
fluctuations on all scales.  Since the $P_{T\delta}$ term contributes with negative
sign during the absorption epoch, $P_{21}$ is reduced.  It appears however that on 
small scales the cross-correlation between density and temperature is small. 
On large scales, the negative sign of the density-Lya cross-correlation can be
seen where the full calculation lies below than that where Lya fluctuations are
ignored.  If we set $P_{T\delta}$ to zero the analytic calculation gives much 
better agreement with the simulation on small scales.  Clearly, improvements to the
simple model of X-ray heating given in \citet{pritchard06} are necessary to 
resolve this problem.
Note also that the
dashed line is slightly higher than the total contribution shown by
the solid line due to the anti-correlation between X-rays and 
\Lyman fluctuations.

At $z \sim 20$, fluctuations in the X-ray heating are not the dominant
source of 21-cm brightness temperature fluctuations, but rather it is
a combination of density perturbations and \Lyman coupling. While
there are general differences at $z \gtrsim 10$ between the analytical model
and our numerical simulation, we note that these still lead to
predictions that agree within a factor of a few, while at $z < 10$,
the agreement is better than 30\% and at $z\sim 7$ it is
significantly better over the wavenumber range where we can make a
comparison. Given that simple fitting formulae can be written down to
model the ionization fraction power spectrum, which we outlined in
Section 2.2, it is likely that one can quickly calculate the 21-cm
power spectrum at low redshifts as a function of the density field
power spectrum. For reasons that we have previously discussed to
enable quick exploration of the parameter space of 21-cm experiments,
we suggest that analytical calculations, complemented by
well-calibrated fitting formulae, can be trusted. For parameter estimation with 
data from upcoming interferometers, it may be necessary to improve beyond
the fitting functions, however.

We note that comparisons such as the one we have performed between an
analytical model and a numerical simulation of reionization, which was
post-processed to extract properties of the 21-cm brightness
temperature, have been performed in the literature and leading to
conclusions similar to ours that analytical calculations are adequate
for parameter predictions and measurements with 21-cm data
\citep{zahn06}. Our work differs from these previous comparisons in
that, for the first time to the extent we are aware of, we extend the
comparisons to redshifts beyond 8 and comment on the agreement even
out to $z \sim 20$.  To do this, we are forced to model fluctuations
in X-ray heating and \Lyman intensity field since as we have shown
these inhomogeneous sources make significant contributions to 21-cm
brightness temperature fluctuations at $z \gtrsim 10$.  At low redshifts,
even after including fluctuations in \Lyman coupling and X-ray
heating, we find that the fluctuations in the 21-cm brightness
temperature are dominated by fluctuations in the ionization fraction
and the density field. Thus, previous analytical and numerical
simulation comparisons, which only considered physics on ionization
bubbles and their distribution in detail, remain valid.

\section{Summary and conclusions}

Using a new large volume, high resolution simulation of cosmic
reionization based on a hybrid code for N-body dark matter and
radiative transfer of ionizing photons through an adaptive algorithm,
we have measured several properties of the reionization process.  We
have focused our discussion on the low-frequency 21-cm signal
associated with the neutral hydrogen distribution which is now pursued
by a variety of interferometers as a probe of the reionization history
of the universe.

In this paper we have studied the extent to which statistical results
from an analytical model are consistent with results extracted from
the simulation. Note that first principle simulations cannot be
considered for baryon physics and star formation so that the
brightness temperature of the simulation 21-cm signal is derived by
post-processing the simulation with certain results based on
analytical prescriptions of the reionization process.

In detail, we have compared the spatial clustering of the neutral gas
fraction, ionization fraction, and the associated 21-cm signal from
the neutral hydrogen distribution.  Our study extends to high
redshifts where the contribution from spin temperature is
non-negligible and we take into account the heating of the gas by
X-rays and the effect of \Lyman and inhomogeneous collisional coupling
when calculating the 21-cm radio signal.  We find very good agreement
between simulations and an analytical model at low redshifts, though
there are, non-negligible differences at higher redshifts ($z \gtrsim 10$)
arising from differences related to X-ray heating and fluctuations in
the \Lyman coupling.  At the redshift range that will be probed by the
first-generation 21-cm experiments ($z < 9$), we find that simple
analytical models coupled with fitting functions associated with the
ionization fraction power spectrum can easily reproduce the results
from the numerical simulation.  Therefore, at these redshifts, we find that there are no
remaining issues with using an analytical model to explore the
parameter space relevant for future
21-cm surveys, when using estimators based on the power spectrum
alone. At higher redshifts, detailed comparisons, especially with
regard to gas temperature and \Lyman coupling, may be desirable
in order to improve current analytical models.

\acknowledgments

This work was partially supported by FCT-Portugal under grant
PTDC/FIS/66825/2006 and was
supported at UC Irvine by NSF CAREER AST-0645427 and by
NASA through grant number 11242 from the Space Telescope Science
Institute, which is operated by the Association of Universities for
Research in Astronomy, Inc., under NASA contract NAS5-26555.  This
research is also supported in part by grants AST-0407176 and
NNG06GI09G.  AA is supported in part by a McCue Fellowship. HT is
supported in part by NASA grant LTSA-03-000-0090.  MGS was partially 
supported by FCT-Portugal under grant BPD/17068/2004/Y6F6.
JRP is supported by NASA through Hubble Fellowship
grant HST-HF-01211.01-A awarded by the Space
Telescope Science Institute, which is operated by the
Association of Universities for Research in Astronomy,
Inc., for NASA, under contract NAS 5-26555.

\begin{appendix}

When non-Gaussian terms become important, equation
\ref{21corr} is no longer valid and we need to take into account the
full 4-point and 3-point function in the power spectrum calculation of
the brightness temperature (see \citealt{lidz06}). The full power
spectrum is then 
\be
\label{21terms}
P_{21}(k)=T_c^2\left[\bar{f}_{\rm HI}^2P_{\delta,\delta}(k)+P_{x_i,x_i}(k)-2\bar{f}_{\rm HI}P_{x_i,\delta}(k)
 +2P_{x_i\delta,x_i}(k)-2\bar{f}_{\rm
    HI}P_{x_i\delta,\delta}(k)+P_{x_i\delta,x_i\delta}(k)\right], 
\ee
where $P_{a,b}$ is just the power spectrum between the quantity $a$
and $b$.

\begin{figure*}[!t]
\centerline{\hspace{-0.5cm}\includegraphics[scale=0.55]{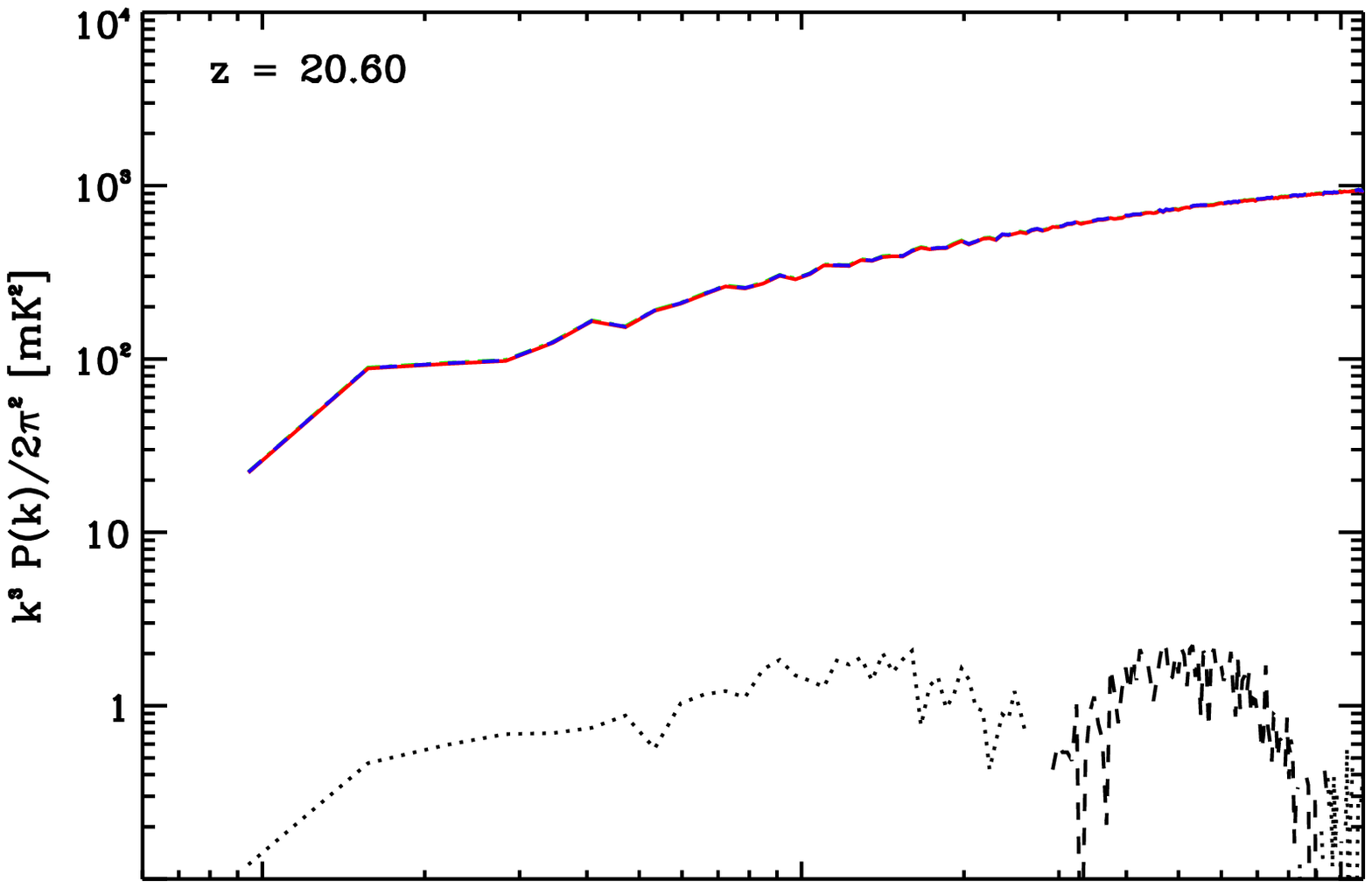}\hspace{-2cm}  \includegraphics[scale=0.55]{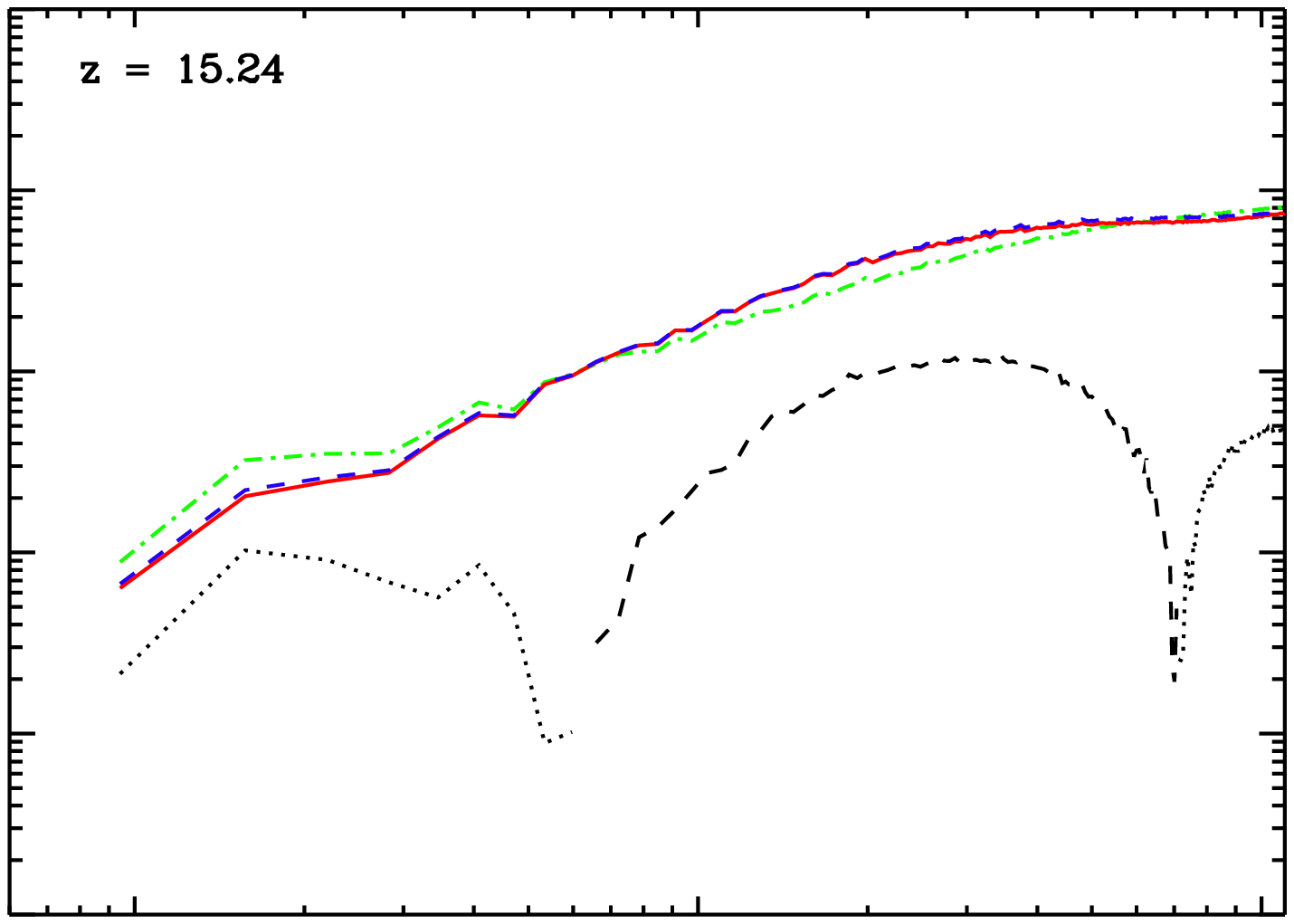}}
\vspace{-1.2cm}
\centerline{\hspace{-0.5cm}\includegraphics[scale=0.55]{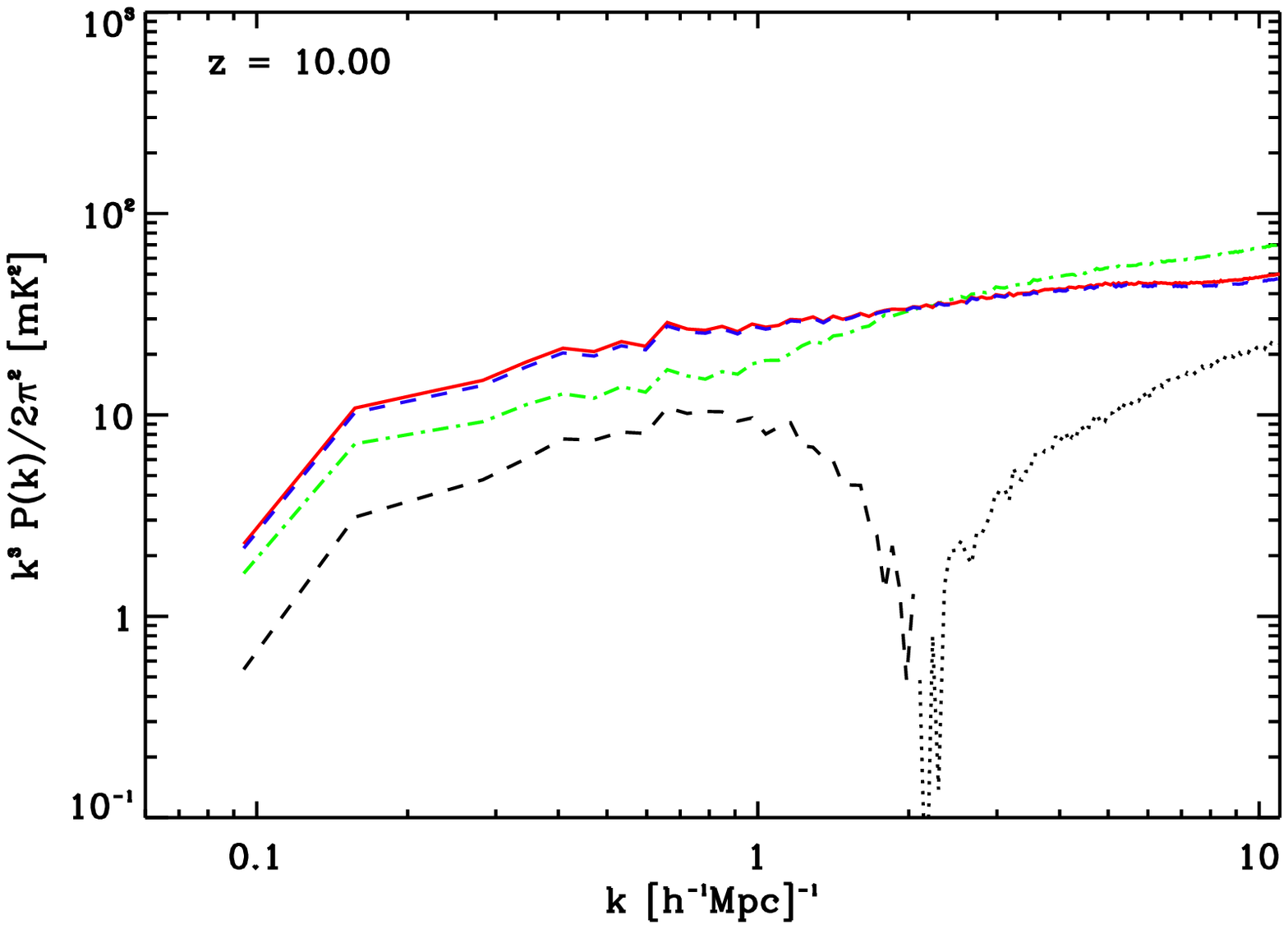}\hspace{-2cm}  \includegraphics[scale=0.55]{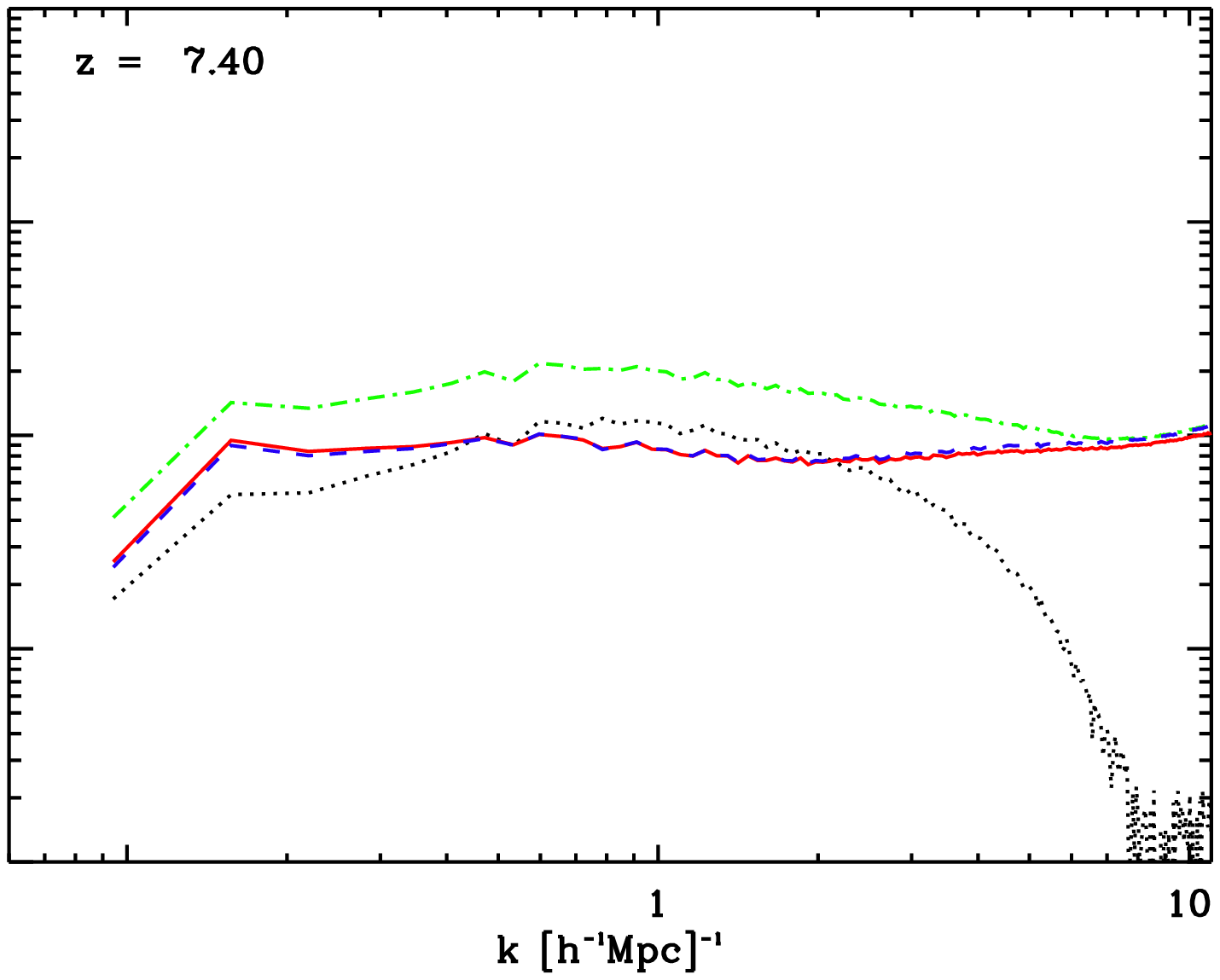}}
\caption{The 3-dimensional power spectra of the brightness temperature
  and the contribution from the low and higher order terms, when only considering
fluctuations in the density and ionization fraction. Red solid
  curves correspond to the power spectrum obtained directly from the
  simulation while blue dashed shows the result of applying equation
  \ref{21terms} which gives similar results as expected. Green
  dot-dashed lines show the contribution from the low order
  terms. Black dotted lines show the negative values of the higher
  order terms while the lower, black dashed line shows the positive
  contribution from the higher order terms. }
\vspace{0.4cm}
\label{T21terms}
\end{figure*}

To see the difference, we show in figure \ref{T21terms} the power
spectrum of the brightness temperature obtained directly from the
simulation (only considering fluctuations in the density and ionization fraction), compared to the one obtained by just using the first three
terms in the equation above (the ``low order'' terms).
Note that all the power spectra used are obtained directly from the
simulation.  We also plot the contribution from the higher order terms
(second line in the equation) plus the result of considering the full
expression, which, as expected, is similar to the actual 21-cm power
spectrum measured from the simulation.

We can see that, only at higher redshifts can the contribution 
of the ``higher order'' terms be safely neglected.  
In addition to accounting for some of the differences in the power spectrum,
non-Gaussianity can also be an important source of information,
specially at the height of the reionization process (i.e., about
$\bar{x_i}\sim 0.5)$, when the ionization structure becomes quite
non-Gaussian.  

\end{appendix}

\end{document}